\documentclass[prd,twocolumn,showpacs,nofootinbib]{revtex4}
\usepackage{graphicx}
\usepackage{amsmath}
\usepackage{bm}
\begin{document}

\newcommand{\Rvir}{$R_{\mathrm{vir}}$}
\newcommand{\Rvire}{R_{\mathrm{vir}}}
\newcommand{\Mvir}{$M_{\mathrm{vir}}$}
\newcommand{\Mvire}{M_{\mathrm{vir}}}
\newcommand{\tdyn}{$t_{\mathrm{dyn}}$}
\newcommand{\tdyne}{t_{\mathrm{dyn}}}
\newcommand{\vk}{$v_{\mathrm{k}}$}
\newcommand{\vke}{v_{\mathrm{k}}}
\newcommand{\vvir}{$v_{\mathrm{vir}}$}
\newcommand{\vvire}{v_{\mathrm{vir}}}
\newcommand{\rhose}{\rho_{\mathrm{s}}}
\newcommand{\rse}{r_{\mathrm{s}}}
\newcommand{\Deltave}{\Delta_{\mathrm{v}}}
\newcommand{\rhoue}{\rho_{\mathrm{u}}}
\newcommand{\rhe}{r_{\mathrm{h}}}
\newcommand{\zre}{z_{\mathrm{re}}}
\newcommand{\vmax}{$v_{\mathrm{max}}$}
\newcommand{\vmaxe}{v_{\mathrm{max}}}
\newcommand{\Modot}{\mathrm{M}_\odot}

\title{Dark-matter decays and Milky Way satellite galaxies}
\author{Annika H.\ G.\ Peter}
\email{apeter@astro.caltech.edu}
\affiliation{California Institute of Technology, Mail Code 249-17, 
  Pasadena, California 91125, USA}
\author{Andrew J. Benson}
\affiliation{California Institute of Technology, Mail Code 350-17, 
  Pasadena, California 91125, USA}

\date{\today}

\begin{abstract}
We consider constraints on a phenomenological dark-matter model consisting of two nearly degenerate particle species using observed properties of the Milky Way satellite galaxy population.  The two parameters of this model, assuming the particle masses are $\gtrsim $ GeV, are \vk, the recoil speed of the daughter particle, and $\tau$, the lifetime of the parent particle.  The satellite constraint that spans the widest range of \vk~is the number of satellites that have a mass within 300 pc $\mathrm{M}_{300} > 5\times 10^6\Modot$, although constraints based on $\mathrm{M}_{300}$ in the classical dwarfs and the overall velocity function are competitive for $\vke \gtrsim 50\hbox{ km s}^{-1}$.  In general, we find that $\tau \lesssim 30$ Gyr is ruled out for $20\hbox{ km s}^{-1} \lesssim \vke \lesssim 200 \hbox{ km s}^{-1}$, although we find that the limits on $\tau$ for fixed \vk~can change constraints by a factor of $\sim 3$ depending on the star-formation histories of the satellites.  We advocate using the distribution of $\mathrm{M}_{300}$ in Milky Way satellites determined by next-generation all-sky surveys and follow-up spectroscopy as a probe of dark-matter properties. 
\end{abstract}

\maketitle

%%%%%%%%%%%%%%%%%%%%%%%%%%%%%%%%%%%%%%%%%%
\section{Introduction}\label{sec:intro}
%%%%%%%%%%%%%%%%%%%%%%%%%%%%%%%%%%%%%%%%%%
Dark matter is the dominant gravitationally attractive component of the Universe \citep{tegmark2004,kessler2009,mantz2010a,vikhlinin2009b,breid2010,larson2010}.  While there is a large set of particle candidates \cite{raffelt1990,turner1990,jungman1996,hogan2000,spergel2000,abazajian2001,cheng2002,feng2003,sigurdson2004,hubisz2005,feng2009}, we have no idea which of these, if any, constitutes the dark matter (although there are some things it clearly \emph{cannot} be, e.g., light neutrinos \cite{white1983}).  However, arguably the most popular candidate class is ``cold dark matter'' (CDM).  This class of candidate, which includes both axions \cite{raffelt1990,turner1990} and weakly-interacting massive particles (WIMPs) \cite{steigman1985}, is called ``cold'' because it is non-relativistic during major events in the early Universe (freeze-out in the case of WIMPs, kinetic decoupling for all).  This class is popular because it is in many ways the simplest; dark-matter candidates come ``for free'' in many extensions to the standard model of particle physics, are in the early Universe at the right abundance in most models, and thereafter evolve in a way that is consistent with observations of large-scale structure.

However, CDM is not the only viable dark-matter candidate class.  A number of observations on smaller scales have inspired investigations into dark-matter models which reproduce the successes of CDM on large scales (corresponding approximately to the scales on which $L_*$ galaxies are observed and larger) while deviating from CDM on the small scales which either lack observations or for which observations are difficult to interpret.  While recent work has focused on finding dark-matter candidates which may boost the light-lepton density throughout the Milky Way  to explain unexpectedly high electron and positron counts \cite{adriani2009,abdo2009,pospelov2009}, the classic arena in which to play the non-cold-dark-matter game is the distribution of dark matter in galaxies.  In particular, the mismatch between the observed number of satellites of the Milky Way and the number of massive subhalos predicted in CDM simulations (coined the ``missing satellites problem'' \cite{moore1999a}) has inspired a number of models in which either the phase-space density of dark matter is reduced or the small-scale power spectrum is cut off (or both) relative to CDM \cite{hogan2000,kaplinghat2005}.

In this paper, we consider a new set of constraints on a class of dark-matter candidate which was originally motivated by the missing-satellites problem and the mass distribution within dwarf galaxies \cite{cen2001,sanchez2003,abdelqader2008}.  This class of model consists of two nearly degenerate massive dark-matter species $X$ and $Y$, where the masses are related by $M_Y =M_X(1-\epsilon)$ with $\epsilon \ll1$.  In the simplest scenario, $X$ decays to $Y$ and a massless particle which need not be a standard-model particle.  If $\epsilon$ is sufficiently small, the $Y$ particle receives a non-relativistic velocity kick $v_k = \epsilon$.  Unlike most decaying-dark-matter models, we consider lifetimes $\tau$ that are comparable to the age of the Universe.  Previous work has shown that $\tau \gtrsim 100$ Gyr to be consistent with cosmic microwave background observations if $v_k$ is relativistic \cite{ichiki2004}, and $\tau \gtrsim 30-40$ Gyr for $v_k \gtrsim 100\hbox{ km s}^{-1}$ in order to remain consistent with the observed galaxy-cluster mass function and the galaxy mass-concentration relation \cite{peter2010a,peter2010c}.  Constraints may be tighter if the massless particle belongs to the standard model \cite{cembranos2007,bell2010}, but are so far lacking for $v_k \lesssim 100\hbox{ km s}^{-1}$ in the case that the massless particle decays to neutrinos or does not interact electromagnetically.

We reexamine this model in light of its original context, the observed population of Milky Way satellite galaxies.  A flurry of work in the past several years has highlighted interesting properties of these galaxies that may shed light on both galaxy evolution on the smallest scales and on the nature of dark matter (e.g., \cite{willman2005,zucker2006,zucker2006b,belokurov2006,belokurov2007,simon2007,walker2007,koch2008,strigari_nat2008,tollerud2008,walker2009,belokurov2010,bullock2010,busha2010,kuzio2010,munoz2010}).  In particular, a number of ultra-faint objects have been found in the Sloan Digital Sky Survey (SDSS) \cite{york2000}, which deep photometric and spectroscopic follow-up have shown to be highly dark-matter-dominated galaxies in the Milky Way halo with a nearly constant amount of dark matter within the inner $\sim $ kpc.  In this work, we show which properties of this galaxy population yield robust constraints to the decay parameter space ($v_k$, $\tau$).  We use a hybrid method combining semi-analytic dark-matter merger trees and star-formation prescriptions with simulations of decay in isolated dark-matter halos to determine the properties of subhalos and satellite galaxies in a decaying-dark-matter cosmology to allow for a comparison with the currently known satellite population.  For the purposes of this work, we define ``subhalo'' as any distinct dark-matter clump within a host dark-matter halo, and ``satellite'' as a subhalo that contains stars.  In addition, we will highlight how the uncertainty in the evolution of baryons in low-mass halo complicates inferences about dark-matter properties from Milky Way satellites.  We show that the resulting constraints on decaying dark matter are complementary to those obtained in Refs. \cite{peter2010a,peter2010c}.

The organization of this paper goes as follows.  In Sec. \ref{sec:methods}, we describe the method by which we constrain decaying dark matter with Milky Way satellites, and in Sec. \ref{sec:results}, we describe the constraints.  In that section, we explore which properties of the Milky Way satellites may robustly constrain the decay parameter space, and show how the constraints depend on the star-formation properties of dwarf galaxies.  In Sec. \ref{sec:discussion}, we place our findings in the context of other work.

%%%%%%%%%%%%%%%%%%%%%%%%%%%%%%%%%%%%%%%%%%
\section{Methods}\label{sec:methods}
%%%%%%%%%%%%%%%%%%%%%%%%%%%%%%%%%%%%%%%%%%
In this section, we describe the key observational properties of Milky Way satellites that we will use to constrain decay properties, and introduce the method by which we calculate the effects of decay on satellite populations of Milky Way-like dark-matter halos.  We describe the observables first because they dictate the requirements for the decay simulations.

\subsection{Observational constraints}\label{sec:methods:obs}
As highlighted in the Introduction, the number of known Milky Way satellites has approximately doubled (with the exact number a moving target) in the past five years due to the advent of sophisticated color-magnitude filtering techniques to find low surface brightness galaxies in the SDSS \cite{willman2005,belokurov2007,walsh2009}.  These galaxies have a number of interesting properties, such as having extremely low luminosities (Segue 1 has $L \approx 300 L_\odot$) and being incredibly dark-matter dominated (again, $M/L\gtrsim 10^3$ within the half-light radii) \cite{geha2009,wolf2010}.  The galaxies are pressure-supported, and have stellar line-of-sight velocity dispersions $\sigma_{\mathrm{LOS}}$ of only a few $\hbox{km s}^{-1}$, significantly smaller on average than the classical Milky Way dwarf galaxies.  To compare these line-of-sight velocities with theoretical predictions for dark-matter halos, $\sigma_{\mathrm{LOS}}$ is often converted to an estimated $\vmaxe =\max (\sqrt{GM(r)/r})$, the maximum circular velocity of the satellite.  Below, we will use the relation used by \citeauthor{madau2008b} \cite{madau2008b}, $\vmaxe = \sqrt{3}\sigma_\mathrm{LOS}$, to compare the \vmax~of subsets of the subhalos in our simulations to those of real galaxies.  Mass-modeling of a subset of classical dwarfs indicates that this relation between $\sigma_{\mathrm{LOS}}$ and \vmax~is reasonable \cite{strigari2010}.

Another interesting, and related, property of the satellites is the inferred mass enclosed within 300 pc of the galaxy centers, $\mathrm{M}_{300}$.  Several analyses have indicated that this mass is nearly constant among the galaxies ($\mathrm{M}_{300}\sim 10^7M_\odot$), even though the luminosities span roughly five orders of magnitude \cite{walker2007,strigari2007c,strigari_nat2008}.  The least luminous spectroscopically confirmed galaxy, Segue 1, has $\mathrm{M}_{300} \gtrsim 5\times 10^6 M_\odot$. Whether this narrow range of $\mathrm{M}_{300}$ is an artifact of the selection function of dwarf galaxies or a fundamental limit in star-formation physics is a matter of debate \cite{bullock2010}, but in any case, it means that the minimum constraint for a dark-matter model is that it produces at least enough subhalos with $\mathrm{M}_{300}$ in the range matching the observed Milky Way satellites, within a Milky Way-mass dark-matter halo.  Overshooting the number of satellites above the $\mathrm{M}_{300}$ threshold is all right because it is generally easier to remove mass from the inner part of a halo than it is to add mass.

The third observed property of these satellites which is relevant to this work is that they have stars.  Although an obvious point, it highlights the fact that any inference about dark matter from the Milky Way satellite population depends on galaxy formation and evolution.  Some authors have estimated the luminosity function for the full population of Milky Way satellites (taking into account SDSS sky coverage and completeness) \cite{koposov2008,tollerud2008,bullock2010}, and used that to test cold or warm dark matter paradigms \cite{maccio2010a,maccio2010b}.  We refrain from using the luminosity function to constrain the decay model because it is highly sensitive to the poorly understood star-formation and feedback processes of small galaxies.  Instead, we simply consider the fact that the observed satellites necessarily contain stars.

In summary, there are three observed or inferred properties of the currently known population of Milky Way satellite galaxies that we use to constrain the decaying dark matter scenario: \vmax, $\mathrm{M}_{300}$, and the presence of stars.  In addition, we know how many satellite galaxies are currently known, and one can estimate the total number of satellite galaxies with properties similar to known satellites by taking into account the sky coverage of SDSS and its selection function \cite{koposov2008,tollerud2008}.  We discuss the details and subtleties of comparing these properties to the simulated properties of satellites in a decaying-dark-matter cosmology in Sec. \ref{sec:results}.

%%%%%%%%%%%%%%%%%%%%%%%%%%%%%%%%%%%%%%%%%
\subsection{Hybrid decay simulator}\label{sec:methods:hybrid}
%%%%%%%%%%%%%%%%%%%%%%%%%%%%%%%%%%%%%%%%%
In order to constrain \vk~and $\tau$, we study the \vmax, $\mathrm{M}_{300}$, and stellar properties of subhalos and satellites for an ensemble of Milky Way-mass halos.  We want to explore an ensemble of Milky Way-mass halos for each point in \vk$-\tau$ space in order to get a sense of how likely or unlikely it is for the observed satellite population to resemble the simulated population.  In light of this goal and the properties of the simulated subhalo and satellite populations we use to compare with observations, we construct a hybrid technique involving both semi-analytic modeling and $N$-body simulations to explore the effects of decay on subhalos and satellite galaxies of Milky Way-mass dark-matter halos.  

The fact that we are interested in long ($\tau \gg 1$ Gyr) decay times and non-relativistic \vk~allows us to use CDM initial conditions for our hybrid decay simulator.  We use merger trees from the {\sc Galacticus} semi-analytic model, which are simulated in $\Lambda$CDM cosmologies, and use prescriptions within {\sc Galacticus} to determine the density profiles of the dark-matter halos and subhalos in the absence of decay \cite{galacticus2010}.  The relevant properties of {\sc Galacticus} for this work are summarized in Sec. \ref{sec:methods:merger}.  We use simulations of decay in isolated dark-matter halos to take into account the effects of decay on individual subhalos in the merger trees, as described further in Sec. \ref{sec:methods:simulations}.  We populate subhalos with stars according to the prescription in Sec. \ref{sec:methods:star}.  

This method is much faster to implement and in many ways more robust than using cosmological $N$-body simulations alone, which at first glance would have been the obvious path to take.  Cosmological $N$-body simulations, even zoomed on a particular host halo, have major disadvantages.  First, even the highest-resolution simulations of Milky Way-mass halos can only probe down to a few hundred parsecs of the center of the main halo potential \cite{power2003,springel2008,diemand2008,stadel2009}.  However, while resolution tests exist for main halo centers, systematic resolution tests on subhalos are lacking.  This is a problem if we want to probe the mass within subhalos on scales comparable to the demonstrated resolution limit on the host halo.  However, in setting the subhalo properties in the merger tree, we do use relations among halo properties that are calibrated on large $N$-body simulations and in ranges of mass or redshift that have not been tested by simulations.  Second, each realization of a Milky Way-mass halo at the present best resolution takes months of supercomputing time to run.  Since we want to explore a range of \vk~ and $\tau$, and to simulate an ensemble of Milky Way-mass halos for each set of \vk~and $\tau$, cosmological simulations are highly impractical.

%%%%%%%%%%%%%%%%%%%%%%%%%%%%%%%%%%%%%%%%%%
\subsubsection{Merger tree}\label{sec:methods:merger}
%%%%%%%%%%%%%%%%%%%%%%%%%%%%%%%%%%%%%%%%%%
Distributions of dark matter subhalo properties at $z=0$ were computed using the {\sc Galacticus} semi-analytic code\footnote{Specifically, \protect\href{http://bazaar.launchpad.net/~abenson/galacticus/v0.9.0/revision/12}{v0.9.0, revision 12} of {\sc Galacticus} was used. The {\sc Galacticus} model can be downloaded from \protect\href{http://sites.google.com/site/galacticusmodel}{http://sites.google.com/site/galacticusmodel}. The input parameter file used for these calculations is available at \protect\href{http://www.ctcp.caltech.edu/galacticus/parameters/parameters_v0.0.1.xml}{http://www.ctcp.caltech.edu/galacticus/parameters/darkMatterDecays.xml}.} \citep{galacticus2010}. Only dark sector physics (dark matter merger tree construction and subhalo orbital decay via dynamical friction) was included in these calculations---all baryonic physics in {\sc Galacticus} was switched off.

Dark matter merger trees were built using the algorithm described by Ref. \cite{cole2000}. Standard values of the accuracy parameters for this algorithm were used as follows:
\begin{description}
 \item [\texttt{[mergerTreeBuildCole2000MergeProbability]}$=0.1$:] The maximum probability for a binary merger allowed in a single time step. This ensures that the probability is kept small, such the the probability for multiple mergers within a single time step is small;
 \item [\texttt{[mergerTreeBuildCole2000AccretionLimit]}$=0.1$:] The maximum fractional change in mass due to sub-resolution accretion allowed in any given time step when building the tree.
\end{description}
Merger trees were resolved down to halos of mass $10^7 M_\odot$. Mass accretion below this scale was treated as smooth accretion and branches were truncated once they fell below this mass.

Branching probabilities in the merger tree were computed using the algorithm of \citet{parkinson_generating_2008}. The parameters $G_0$, $\gamma_1$ and $\gamma_2$ of their algorithm were set to $0.57$, $0.38$ and $-0.01$ respectively as recommended by \citet{parkinson_generating_2008}. Additionally, the parameter \texttt{[modifiedPressSchechterFirstOrderAccuracy]} in {\sc Galacticus} was set to $0.1$ to limit the step taken in the critical linear theory overdensity for collapse in the merger tree building algorithm. This step was not allowed to exceed \texttt{[modifiedPressSchechterFirstOrderAccuracy]} times $\sqrt{2[\sigma^2(M_2/2)-\sigma^2(M_2)]}$, where $M_2$ is the mass of the halo being considered for branching and $\sigma(M)$ is the CDM mass variance computed by filtering the power spectrum using top-hat spheres. This ensures that the first order expansion of the merging rate that is assumed in the tree building algorithm is accurate.

Progenitor halo mass functions from merger trees built using this algorithm have been compared with equivalent progenitor mass functions measured from the Millennium Simulation \citep{springel2005} and show excellent agreement with the N-body result \cite{parkinson_generating_2008,galacticus2010}.

The {\sc Galacticus} code evolves the merging distribution of halos forward in time. When one halo merges with another, larger halo it becomes a subhalo within that larger host. We track only a single level hierarchy of substructure, i.e. we track only substructures, not sub-substructures or deeper levels or the merging hierarchy. Therefore, if a merging halo contains its own subhalos they will become independent subhalos within the new host and will be assigned new merging times (see below).

Once a halo becomes a subhalo it is assigned a timescale for merging to the center of its host halo due to the actions of dynamical friction. We use the dynamical friction calibration of \citet{jiang2008} to compute dynamical friction timescales, with orbital parameters of subhalos selected at random from the cosmological distribution found by \citet{benson_orbital_2005}. Once this timescale for merging has elapsed the subhalo is merged into its host and no longer exists as an independent entity.

The properties of the subhalos, including the density profile and the mass, are set at accretion time.  The radius at which the slope of the density profile $\rse = d \log \rho / d \log r = -2$ relative to the virial radius \Rvir, defining the concentration parameter
\begin{equation}
c = \Rvire/\rse
\end{equation}
is set according to \citet{gao2008}, which is calibrated using the Millennium Simulation \cite{springel2005}.  The subhalo mass and virial radius at accretion are set according to the virial overdensity criterion of \citet{percival2005} for homogeneous dark-energy CDM cosmologies.

There are a few caveats to applying these particular options for the merger tree.  First, the merger tree and, for example, the mass-concentration relation as a function of redshift have been tested on a small set of simulations representing a limited set of cosmological parameters and range of halo mass.  For example, the redshift-dependent concentration and the merger histories have been calibrated using the Millennium Simulation, which has a relatively high $\sigma_8 = 0.9$, and for which the dark-matter particle mass is $\sim 10^9\Modot$ (larger than many of the subhalos in our merger trees) \cite{springel2005,gao2008}.  Others have found that the mass-concentration relation depends on a number of cosmological parameters, in particular $\sigma_8$, and the redshift-evolution of this relation is still under debate \cite{bullock2001,huffenberger2003,maccio2008}.  Moreover, this relation has not been tested in the mass and redshift ranges of some of the subhalos before they merge onto a larger halo.  The main way we mitigate some of these uncertainties is to impose a minimum cut-off in the concentration.  Studies have shown that $c\approx 4$ at virialization, so we assign high-redshift halos $c=4$ if the \citet{gao2008} formula indicates $c$ below that value \cite{zhao2009}.  In addition, there is a great deal of scatter in the mass-concentration even in the mass and redshift range in which it has been studied.  The dynamical friction formula has been determined using a set of N-body cosmological simulations, and also not necessarily on the small and early scales that are relevant to this work \cite{jiang2008}.

Finally, the merger tree does not include the effects of tidal stripping.  Tidal stripping not only destroys some subhalos, but also may reduce \vmax~and $\mathrm{M}_{300}$ of the remaining subhalos.  The important implication of this fact for this work is that our constraints are quite conservative, as we use the subhalo properties at the time of accretion.

%%%%%%%%%%%%%%%%%%%%%%%%%%%%%%%%%%%%%%%%%%
\subsubsection{Decay simulations}\label{sec:methods:simulations}
%%%%%%%%%%%%%%%%%%%%%%%%%%%%%%%%%%%%%%%%%%
In order to estimate the effects of decay on the subhalo population and the host halos, we use a set of simulations of isolated, initially equilibrium CDM halos.  The first set of simulations we use, with $c = 5$ and $10$, were initially presented in \citet{peter2010c}.  The halos in these simulations had initial virial mass $\Mvire = 10^{12} \Modot$, with the virial overdensity defined in Ref. \cite{bryan1998}.  There were 25 sets of simulations, scanning decay parameters $\vke/\vvire = 0.077, 0.38,0.77,1.54$ and $3.85$, and $\tau = 0.1, 1, 10, 50,$ and 100 Gyr, where \vvir~is the virial velocity of the halo.  For this work, we simulated $\Mvire = 10^{12}\Modot$ halos with one million particles each of mass $10^6 \Modot$ with $c=20$ and $c=30$ for the same sets of $\vke/\vvire$ and $\tau$ in order to span a broad range of subhalo concentrations.  We simulated these halos using a modified version of Gadget-2 \cite{springel2005}.  As in \citet{peter2010c}, we assumed a Navarro-Frenk-White (NFW) density profile for the initial matter distribution within the virial radius \Rvir~of the halo \cite{navarro1996,navarro1997}, 
\begin{eqnarray}\label{eq:rho}
	\rho(r) = \frac{\rho_\mathrm{s}}{\displaystyle \frac{r}{r_\mathrm{s}}\left(1+\frac{r}{r_\mathrm{s}}\right)^{2}},
\end{eqnarray}
where $\rho_\mathrm{s}$ is the scale density, and $r_\mathrm{s}$ is the scale radius.  At the level of numerical resolution in our simulations, this density profile is nearly indistinguishable from the now-preferred Einasto profile \cite{navarro2010}.  All other aspects of the simulation, from the set-up of the initial conditions to the simulation and cosmological parameters, is identical to that in \citet{peter2010c}.

To apply these simulations to the subhalos of the merger trees, as well as to the host halos, we use the following strategy.  To characterize the subhalos, we use $\rse$ and $\rhose$ at the time of accretion, and find the mass and concentration of these subhalos at $z=0$ assuming that the only growth to the subhalos comes from the decrease in the global mass density of the Universe as a function of time.  This sets the $z=0$ concentration parameter and virial mass, and we estimate the effects of decay using these properties.  We characterize the isolated decay simulations by five parameters: $\vke/\vvire$, which relates the recoil speed to the typical speed of particles in the halo (and hence, the escape velocity); $c$, which characterizes the depth of the potential well in addition to the typical dynamical time scale of particles within the halo (since the dynamical time at the half-mass radius depends only on $c$); the decay time scale $\tau$; $r/\Rvire$, the radius at which we sample the mass profile of the halo as a function of the virial radius; and time $t$.  We divide the mass profile by the initial virial mass in the simulation, and we cut off the mass profile at $r = 0.04\Rvire$ since we have found that numerical relaxation becomes a problem for smaller radii.  For a given subhalo with specific $z=0$ properties, we find the subhalo mass profile by interpolating the simulations in the log-space of $\vke/\vvire$, $c$, $\tau$, and $t$.  We use the mass profile to determine $\mathrm{M}_{300}$ and \vmax~ at $z=0$, and \vmax~at the accretion time (since this determines whether or not the subhalo has stars, as will be described below).  

We also apply the decay interpolation to the host halos, since decays can obviously change the mass of the host as well as the subhalos.  This is important for the host mass cut we employ in Sec. \ref{sec:results}.

This provides a conservative estimate of the effects of decay on the subhalos since the interplay between decay and tidal forces are likely to reduce \vmax~and $\mathrm{M}_{300}$ even more than if the subhalos were isolated halos.  

%%%%%%%%%%%%%%%%%%%%%%%%%%%%%%%%%%%%%%%%%%%
\subsubsection{Populating subhalos with stars}\label{sec:methods:star}
%%%%%%%%%%%%%%%%%%%%%%%%%%%%%%%%%%%%%%%%%%%
Since we only care if a subhalo has stars at all, a crude model for populating halos with stars is acceptable.  If a halo is accreted onto a larger halo prior to reionization, the halo is allowed to have stars if its maximum circular velocity at accretion $\vmaxe > 2\hbox{ km s}^{-1}$, which is approximately the lowest threshold at which gas may accrete onto halos and cool via collisional interactions with molecular hydrogen \citep{haiman1996a,tegmark1997,wise2007}.  During and after reionization, star formation is suppressed in low-mass halos due to various effects related to the strong ionizing background radiation \citep{thoul1996,barkana1999,benson2002a,benson2003,dijkstra2004}.  To model these effects, we allow halos with $\vmaxe > 38 \hbox{ km s}^{-1}$ at accretion onto a larger halo after reionization to host stars; those with smaller \vmax~are not allowed to host stars.  Though this step-function treatment of stellar content in halos is crude, it captures the essence of the fact that star formation in halos depends on reionization and that the stellar content of a subhalo depends on its accretion history.  This model for populating halos with stars is similar to that adopted by \citet{madau2008b}.

\section{Results}\label{sec:results}
For each set of (\vk, $\tau$), we selected $\sim 100$ host halos with virial masses in the range $(0.5-2)\times 10^{12}\Modot$, since this appears to be the plausible range of mass for the Milky Way halo (although most estimates favor the higher end of this range) \cite{zaritsky1989,wilkinson1999,sakamoto2003,battaglia2005,battaglia2006,dehnen2006,li2008,xue2008,gnedin2010,mcmillan2010,watkins2010}.  We have checked the results in this section for dependence on the merger history.  Specifically, simulations suggest that the Milky Way disk could not have withstood a 10:1 merger since $z\sim 1$ \cite{stewart2008,purcell2009a}.  We found that the results from the entire host-halo population were indistinguishable from those from the sample of hosts selected to have not had a major merger since $z=1$.
 
In this study, we consider several subsets of the subhalo populations.  The first sample we call the ``all nodes'' sample, as it represents all nodes of the merger tree down to the mass resolution of $10^7 \Modot$.  This is the most conservative subhalo sample to use to constrain $\vke-\tau$ parameter space because we ignore dynamical friction and we characterize the subhalos by their properties at the moment they are accreted onto larger halos.  The second subset of subhalos we use is the ``dynamical friction'' sample.  This contains all subhalos in the merger tree that do not sink to the center of larger halos by dynamical friction.  The survival probability is calculated according to Ref. \cite{jiang2008} and is the default setting of {\sc Galacticus} \cite{galacticus2010}. The mass profile of these subhalos varies continuously prior to when they first become a subhalo (in either the main branch of the merger tree or a sub-branch). After that time, their mass profile remains fixed. Using this sample to set constraints is less conservative than using the ``all nodes'' sample because it is not completely clear how decays will affect dynamical friction; on one hand, decays decrease the mass of both the subhalo and host (although the effect on the former is stronger than on the latter), reducing the effectiveness of dynamical friction; on the other hand, the binding energy of the subhalo decreases, making it easier to shred (although this effect is not modeled in {\sc Galacticus} for CDM halos).

\begin{figure*}[t]
\centering
\includegraphics[width=0.49\textwidth]{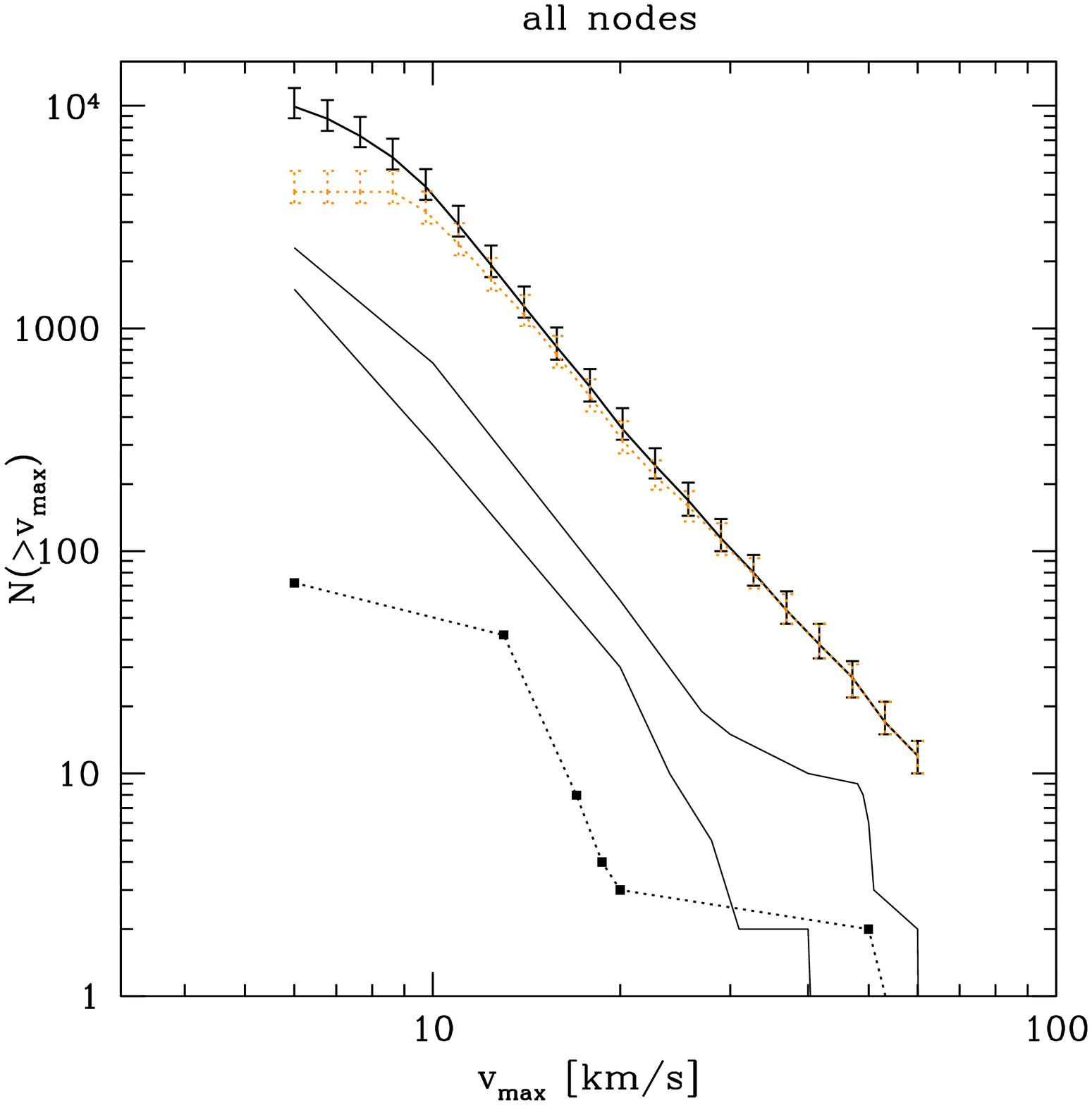}
\includegraphics[width=0.49\textwidth]{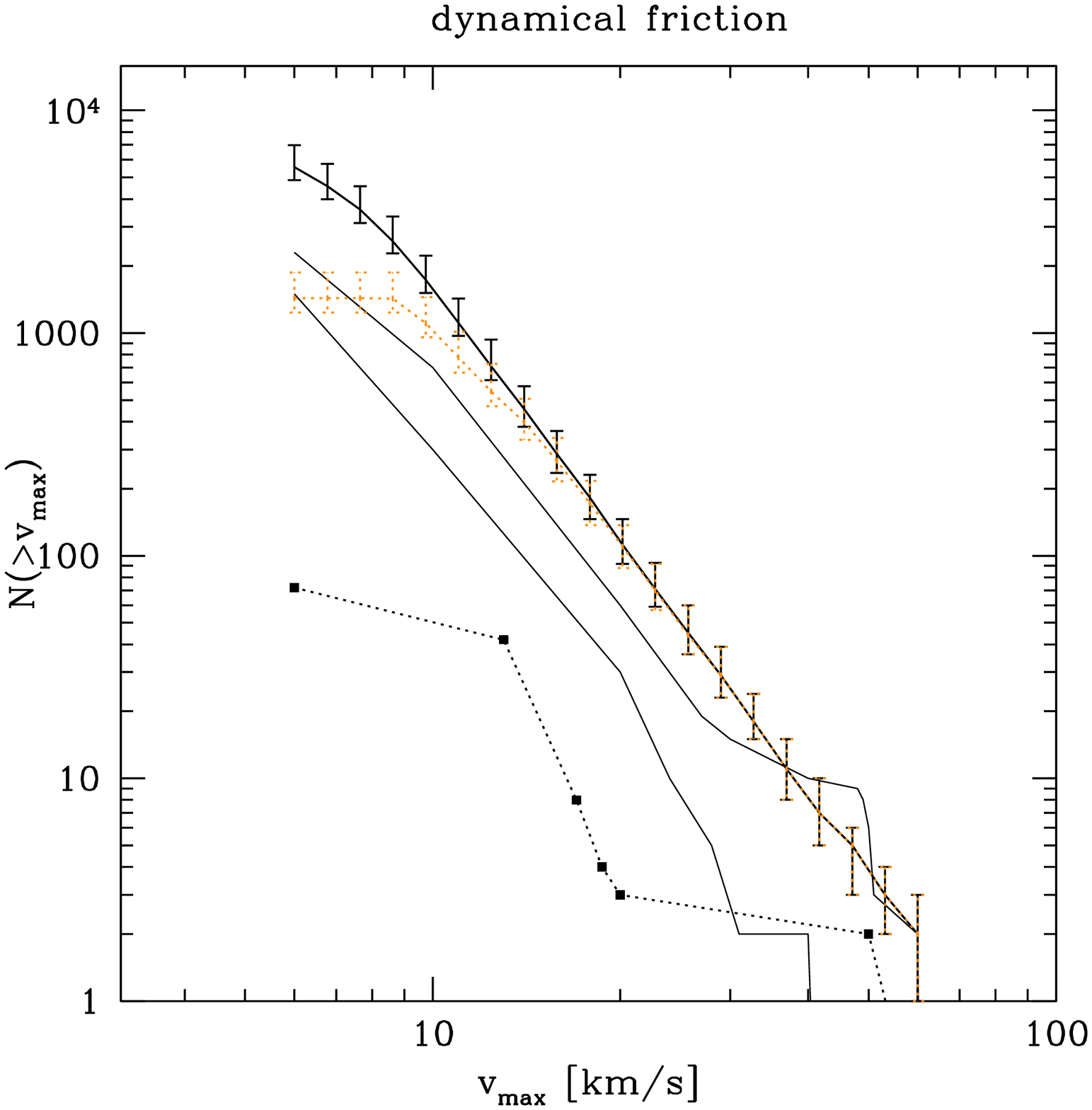}
\caption{\label{fig:vhist_cdm}Velocity function of subhalos.  The thick solid lines with error bars represent the CDM velocity functions from the merger trees for the ``all nodes'' (left) and ``dynamical friction'' (right) samples.  The dotted line with error bars show the velocity function for subhalos with $\mathrm{M}_{300} > 5\times 10^6\Modot$.  The dotted line connecting data points represents an estimated SDSS sky coverage-corrected velocity function for known Milky Way satellites \cite{madau2008b}, and the thin lines (upper: Aquarius A \cite{springel2008}; lower: Via Lactea II \cite{madau2008b}) represent velocity functions found in high-resolution CDM simulations.}
\end{figure*}

The differences in the CDM properties of these two populations is shown in Fig. \ref{fig:vhist_cdm}.  We show these properties for CDM because they will influence the properties of the ``all nodes'' and ``dynamical friction'' populations once we turn on the decays.  What we show in this figure is the velocity function, the number of subhalos above a maximum circular velocity \vmax, $N(>\vmaxe)$, for both the ``all nodes'' sample (left) and ``dynamical friction'' sample, with the error bars showing the 25\% and 75\% percentiles of the number of subhalos in the host halos.  The percentiles should be taken as an indication of the range of \vmax~and not interpreted strictly as errors because we have chosen an initial host population with masses drawn uniformly from the range $1$--$3\times10^{12}M_\odot$ instead of choosing the host masses according to a probability distribution from existing data on the Milky Way host mass.  These merger trees were generated for a flat $\Lambda$CDM cosmology with $\sigma_8 = 0.9$ and $\Omega_\mathrm{m} = 0.2725$ and $n_\mathrm{s} = 0.961$.  Note that since the velocity function is a cumulative function, the percentile bars are highly correlated.  Also on the plots are the velocity function of known Milky Way satellites, corrected for SDSS sky coverage (dotted line connecting data points) \cite{diemand2008}, and the velocity function from two different high-resolution CDM simulations of Milky Way-mass dark-matter halos (thin solid lines).  The upper thin line represents the velocity function found in the high-resolution Aquarius A simulation of a $\Mvire \sim 2\times 10^{12}\Modot$ halo with $\Omega_{\mathrm{m}} = 0.25$, $n_s = 1$, $\sigma_8 = 0.9$ cosmology \cite{springel2008}.  The lower thin line shows the velocity function of the Via Lactea II halo, which has a similar virial mass as the Aquarius A halo but is simulated in a cosmology with $\Omega_\mathrm{m} = 0.238$, $n_s = 0.951$, $\sigma_8 = 0.74$ \cite{madau2008b}.  This velocity function has a lower normalization because the cosmological parameters used tend to produce fewer and less dense subhalos than the ones employed by the Aquarius collaboration.  The velocity functions in the CDM simulations lie below those of our merger trees because the simulations necessarily take into account tidal stripping.  Tidal stripping tends to lower \vmax~after accretion, and \citet{madau2008b} shows that the velocity function of \vmax~at accretion has approximately a factor of five higher normalization than the velocity function for which stripping is taken into account, for $\vmaxe \gtrsim 6\hbox{ km s}^{-1}$.

It is apparent that velocity function of the ``dynamical friction'' sample has both an overall normalization lower than that of the ``all nodes sample'', and that the slope is substantially steeper.  This is to be expected, as dynamical friction is more efficient for higher-mass (and hence, in general, higher \vmax) subhalos.  Thus, in the case of CDM, we find that the ``all nodes'' sample contains both more and on average more massive subhalos.  Moreover, given the factor of $\sim 5$ difference in normalization for the velocity function for \vmax~at accretion versus \vmax~at $z=0$ with tidal stripping included, the ``dynamical friction'' sample is a somewhat better match to the velocity function of the two high-resolution CDM simulations, taking into account differences in underlying cosmologies and host halo masses \cite{springel2008,diemand2008}.

Within each of the ``all nodes'' and ``dynamical friction'' subhalo samples, we determine two different possible satellite populations: a $\zre = 7$ satellite sample consisting of all subhalos satisfying the star-formation criterion for $\zre = 7$ given in Sec. \ref{sec:methods:star}, and a $\zre = 11$ satellite sample consisting of subhalos satisfying that criterion at a reionization redshift $\zre = 11$.  We choose the $\zre = 11$ reionization redshift because WMAP seven-year data suggest that the Universe was reionized at $\zre = 10.5 \pm 1.2$ \cite{komatsu2010}.  However, reionization is ultimately a time- and location-dependent process.  Other work has indicated that the Milky Way-associated region could have been reionized as late as $\zre = 7$ \cite{alvarez2009,busha2010}.  

\subsection{Number of subhalos and satellites above $\mathrm{M}_{300}=5\times 10^6\Modot$}\label{sec:results:number}

Our first constraint on \vk~and $\tau$ comes from considering the number of simulated satellites above an $\mathrm{M}_{300}$ threshold $\mathrm{M}_{300} = 5\times 10^6 \Modot$.  This threshold for observed satellites is apparent in \citet{strigari_nat2008}.  To estimate the true number of Milky Way satellites above this threshold from the observed satellite population, we extend the work of \citeauthor{tollerud2008} \cite{tollerud2008}.  In that work, the authors use the sky coverage and selection function of SDSS as well as the subhalo distribution in the Via Lactea simulation \cite{diemand2007} to estimate the number of satellite galaxies out to the virial radius ($\sim 389$ kpc for the Via Lactea halo, $\Mvire = 1.8\times 10^{12}\Modot$).  They found that there should be, on average, 382 satellites n the Milky Way within that virial radius, and a 98\% probability that there would be at least 292.  Since we consider the possibility that the Milky Way halo could be up to a factor of four less massive (remaining consistent with published estimates of the Milky Way halo mass), we must adjust the \citet{tollerud2008} results to find the minimum number of satellites within the Milky Way virial radius.

We do this by considering the radial distribution of the Via Lactea subhalos, rescaling the distribution by the virial radius.  We find that for possible Milky Way halo masses $\Mvire \gtrsim 5\times 10^{11}\Modot$, we expect at least $\sim 200$ satellites within the Milky Way virial radius.  This lower bound is somewhat rough, as we do not do a full recalculation of \citeauthor{tollerud2008} but instead estimate the average difference in the parameter $f(>r)$ (defined in \citet{tollerud2008}) between a virial mass $5\times10^{11}\Modot$ and $1.8\times 10^{12}\Modot$.  As long as the radial distribution of subhalos is relatively insensitive to host halo mass, and as long as the radial distribution is not a strong function of decay, this approach should yield an approximately correct estimated minimum number of satellites.

In order to set conservative limits on $\vke-\tau$ space, we create merger trees with $\sigma_8 = 0.9$, which is slightly above the 2$-\sigma$ upper limit from the WMAP seven-year data set \cite{komatsu2010}.  We choose a high $\sigma_8$ because structures form earlier and have higher densities for high $\sigma_8$.  The consequences for our study is that there are higher numbers of subhalos per host halo, and those subhalos have higher concentration (and are thus less prone to disruption and will tend to have high \vmax~and $\mathrm{M}_{300}$) for high $\sigma_8$.  In addition, for low-redshift measurements (including the galaxy power spectrum), decays can masquerade as low $\sigma_8$ \cite{peter2010a}.  Although high- and low-redshift estimates of $\sigma_8$ are largely consistent with each other, we choose a high $\sigma_8$ for our study to be conservative \cite{tegmark2004,tegmark2006,vikhlinin2009b,mantz2010a,breid2010,rozo2010}.  We find that the number of subhalos and satellites for $\sigma_8$ at its mean WMAP seven-year values is up to a factor of two less than for the $\sigma_8 = 0.9$ samples.  

\begin{figure}[t]
\centering
\includegraphics[width=0.49\textwidth]{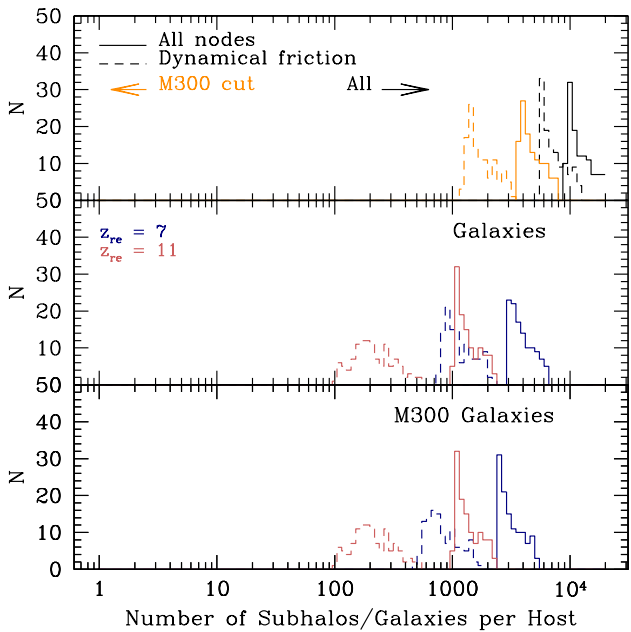}
\caption{\label{fig:m300cdm}Distribution of the number of subhalos or satellites per halo in the host halo sample for CDM.  The upper panel shows the numbers of subhalos in the ``all nodes'' and ``dynamical friction'' samples both with and without the $\mathrm{M}_{300} > 5\times 10^6\hbox{ M}_\odot$ cut.  The middle panel shows the distribution in the number of subhalos satisfying the star-formation criterion for $\zre = 7$ and $\zre = 11$, and the bottom panel shows the distribution in the number of satellites satisfying both the $\mathrm{M}_{300}$ and star formation criteria.}
\end{figure}

In order to compare to decaying-dark-matter cosmologies, in Fig. \ref{fig:m300cdm}, we show the probability distributions for the numbers of subhalos and satellites of the ``all nodes'' and ``dynamical friction'' samples for CDM, according the the merger trees.  Again, the distributions are not actual probability distributions for the Milky Way, as we have not weighted the host halo mass distribution according to the probability distribution of the Milky Way mass from observations.  The distribution is meant to give a sense of the range of possible numbers of subhalos and satellites within the Milky Way halo.  The top panel shows the distribution of subhalos for each sample, and the distribution in the number of subhalos satisfying the $\mathrm{M}_{300} > 5\times 10^6 \Modot$ criterion.  We find that there should be thousands of subhalos satisfying this criterion for either the ``all nodes'' or ``dynamical friction'' samples, with a factor of $\sim 3$ more expected in the ``all nodes'' than ``dynamical friction'' samples.  The middle panel shows the number of subhalos satisfying the star-formation criterion of Sec. \ref{sec:methods:star}, regardless of $\mathrm{M}_{300}$.  There are proportionally far fewer satellites for the ``dynamical friction'' samples than the ``all nodes'' samples because many of the high-\vmax~subhalos in the ``all nodes'' sample have merged with host via dynamical friction.  The lowest panel shows the distribution in the number of subhalos satisfying both the $\mathrm{M}_{300}$ and star-formation criteria.  We find little in the way of differences between the middle and lowest panels of Fig. \ref{fig:m300cdm} because of a mild correlation between $\mathrm{M}_{300}$ and \vmax, and a strong correlation between halo formation time and $\mathrm{M}_{300}$, such that high-\vmax~halos that can form stars after reionization also have high $\mathrm{M}_{300}$, and the medium-\vmax~halos that form stars prior to reionization form early and thus have high $\mathrm{M}_{300}$.

\begin{figure*}[t]
\centering
\includegraphics[width=0.49\textwidth]{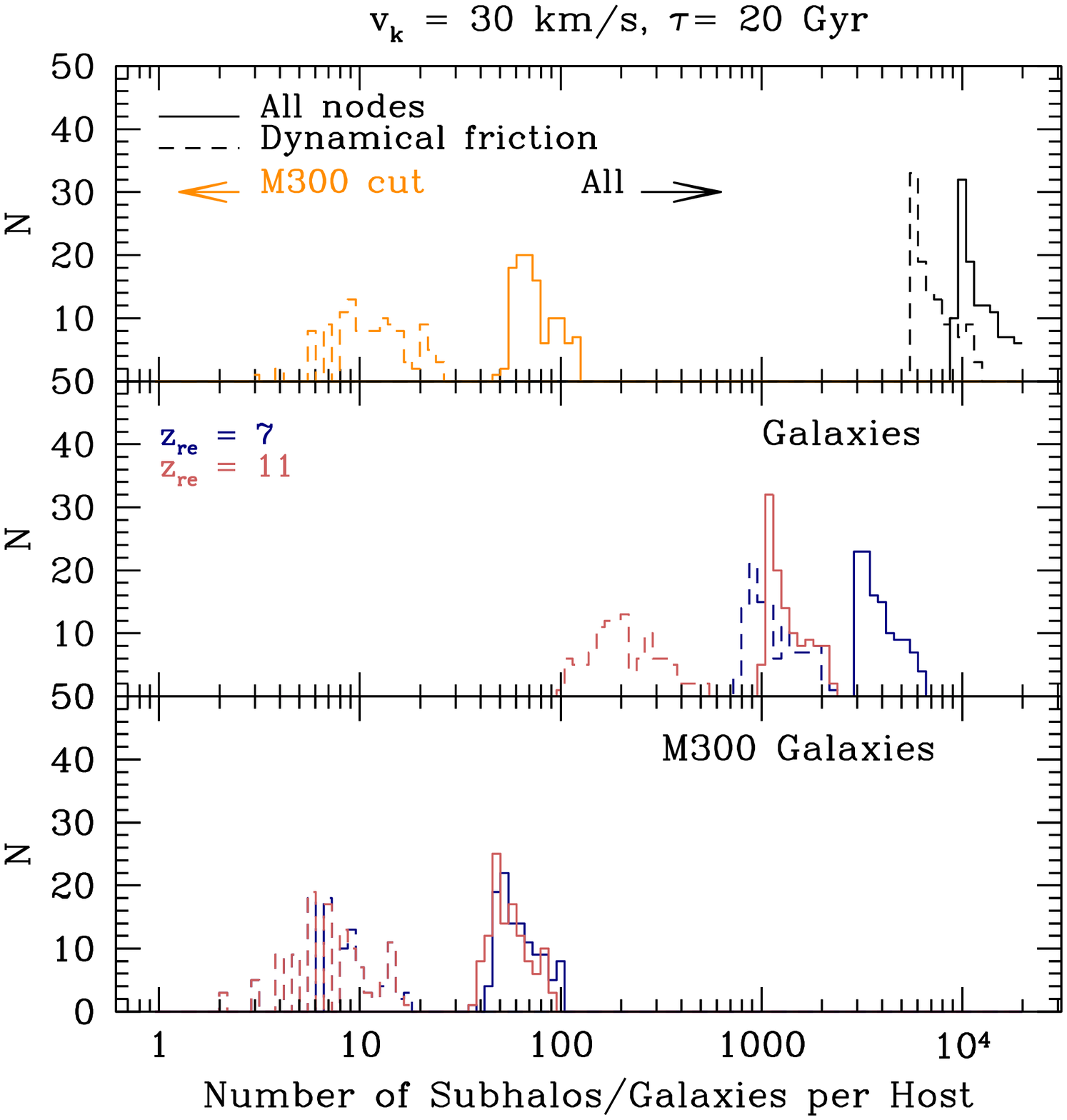}
\includegraphics[width=0.49\textwidth]{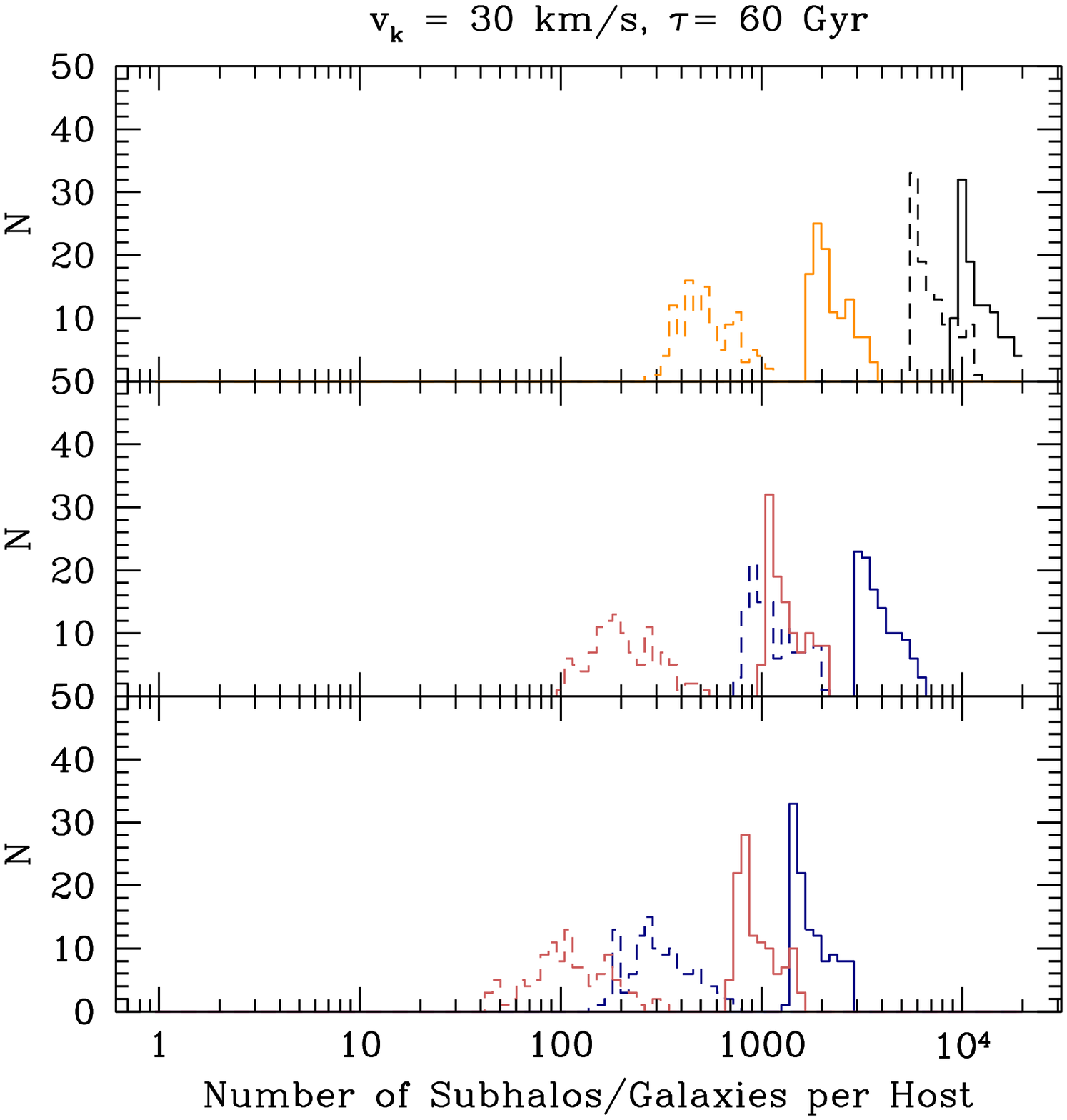}\\
\includegraphics[width=0.49\textwidth]{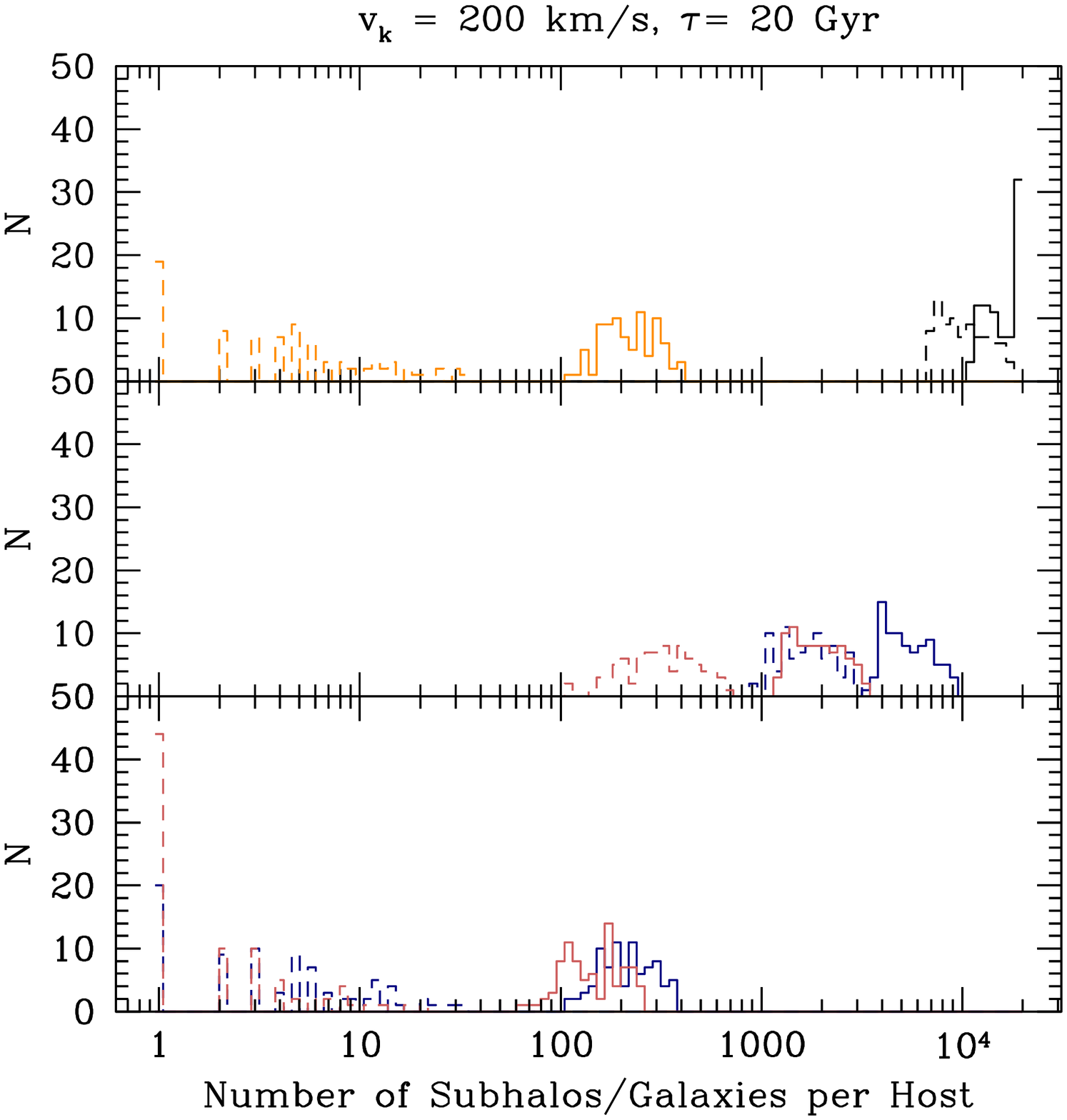}
\includegraphics[width=0.49\textwidth]{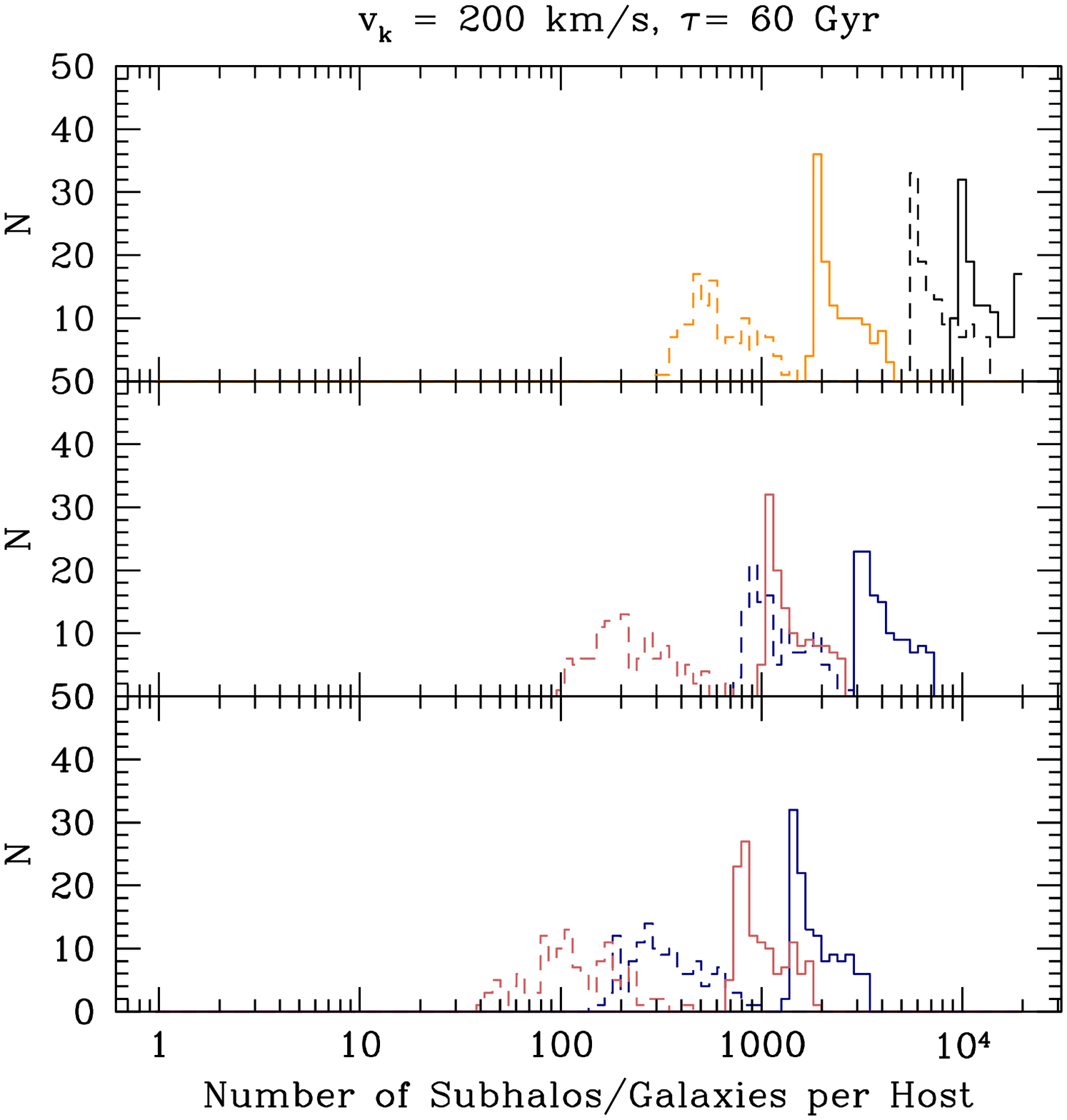}
\caption{\label{fig:m300comp}Distribution of the number of subhalos or satellites per halo in the host halo sample.  Each plot shows distributions for fixed \vk~and $\tau$ (\emph{top row:} $\vke = 30\hbox{ km s}^{-1}$; \emph{bottom row:} $\vke = 200\hbox{ km s}^{-1}$; \emph{left column:} $\tau = 20\hbox{ Gyr}$; \emph{right column:} $\tau = 60 \hbox{ Gyr}$).  Panels have the same meaning as in Fig. \ref{fig:m300cdm}.}
\end{figure*}

We illustrate the effects of decay on the subhalo and satellite samples for $\vke = 30$ and $200 \hbox{ km s}^{-1}$ and $\tau = 20$ and 60 Gyr in Fig. \ref{fig:m300comp}, which gives a flavor of what decay does to the subhalo and satellite populations.  When $\vke = 30\hbox{ km s}^{-1}$, the smaller subhalos tend to be disproportionally affected, since \vvir~or \vmax~of the larger subhalos are a bit bigger than \vk.  Much of the dark matter in the small halos is quickly ejected, and the daughter particles that remain in the halo are responsible for a fairly large (but $\tau$-dependent) injection of kinetic energy, which tends to drastically reduce the central density.  The effect is more pronounced for smaller $\tau$ because the central density is extremely sensitive to the decay fraction if many or most of the daughter dark-matter particles are ejected from the halo after the decay.  The reason that the number of $\zre = 7$ and $\zre = 11$ satellites is nearly identical for $\vke = 30 \hbox{ km s}^{-1}$ is that only the highest-\vmax~halos are largely unaffected by the decays.  

For the same $\tau$, there are more high-$\mathrm{M}_{300}$ subhalos for higher \vk~because the decays start affecting the host halos, too.  If host halos suffer mass loss due to decay, then in order for the host halo to be in our specified range, it must have had a higher CDM halo mass.  Since the number of substructures above a mass threshold is correlated with host mass, the hosts at $z=0$ that are significantly affected by decay and whose $z=0$ mass lies within our target range have more subhalos than if the effects of decay were minimal.  This effect is noticeable if one compares the two upper plots with the two lower plots in Fig. \ref{fig:m300comp}.

\begin{figure*}[t]
\centering
\includegraphics[width=0.32\textwidth]{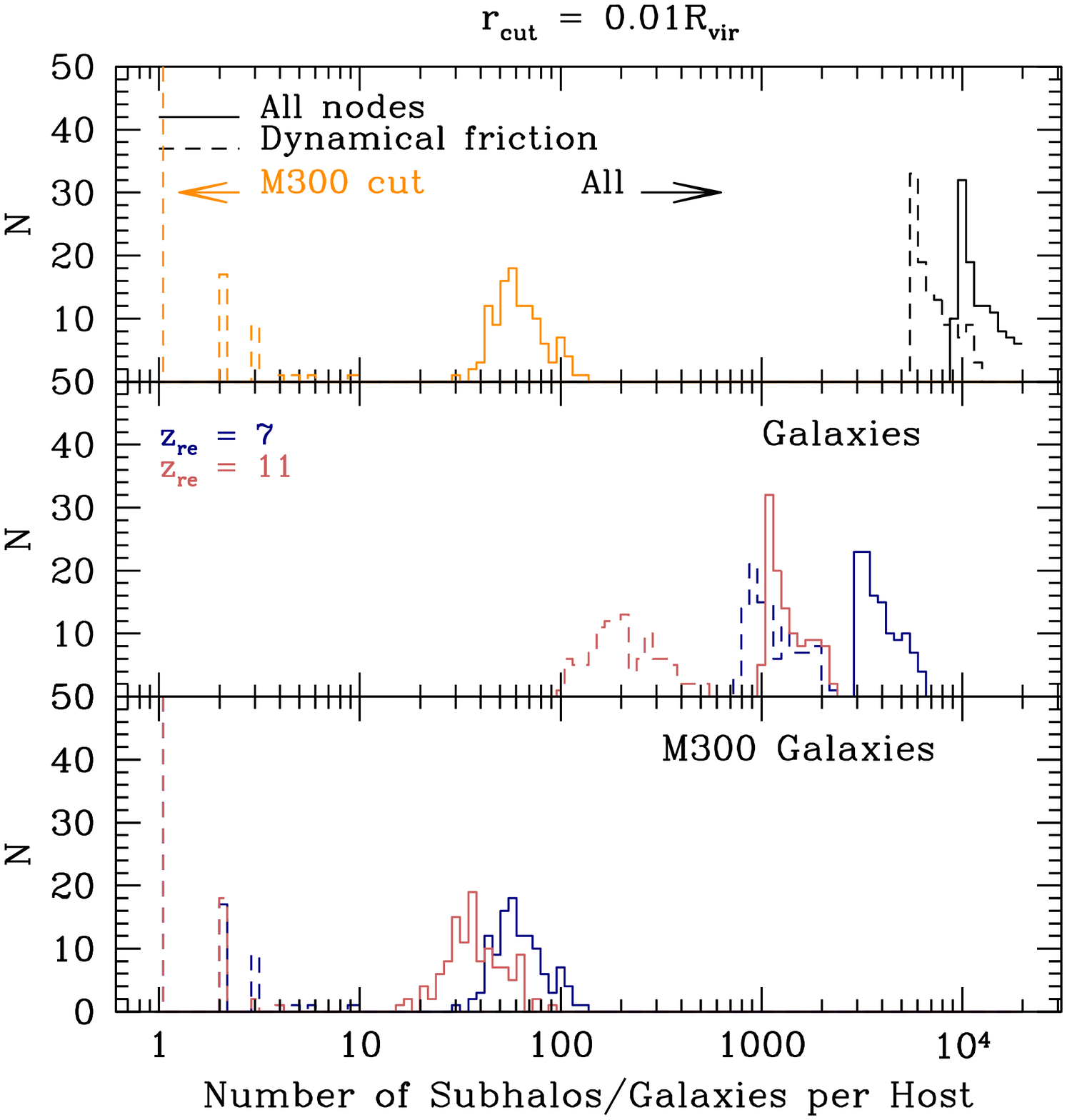}
\includegraphics[width=0.32\textwidth]{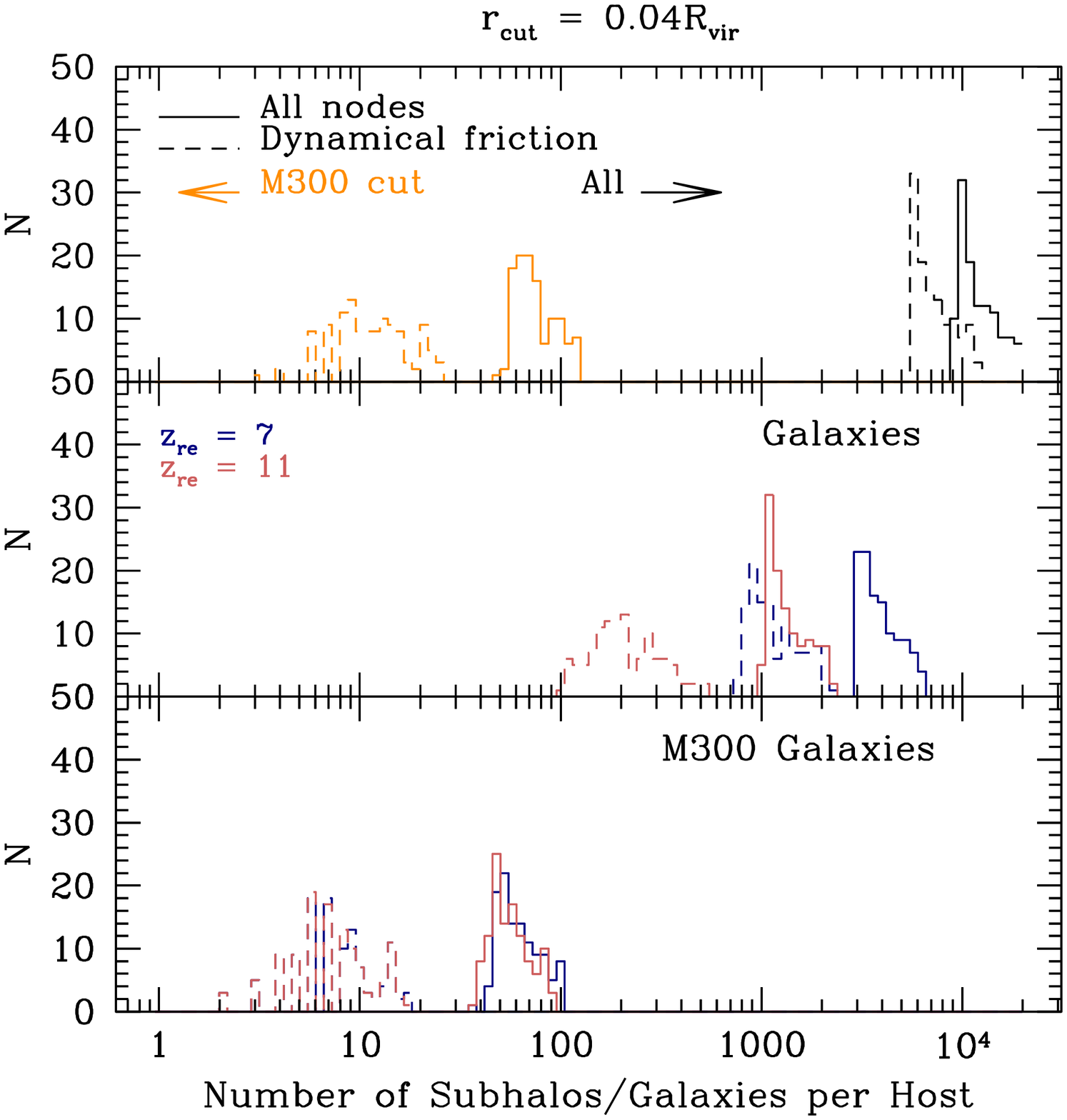}
\includegraphics[width=0.32\textwidth]{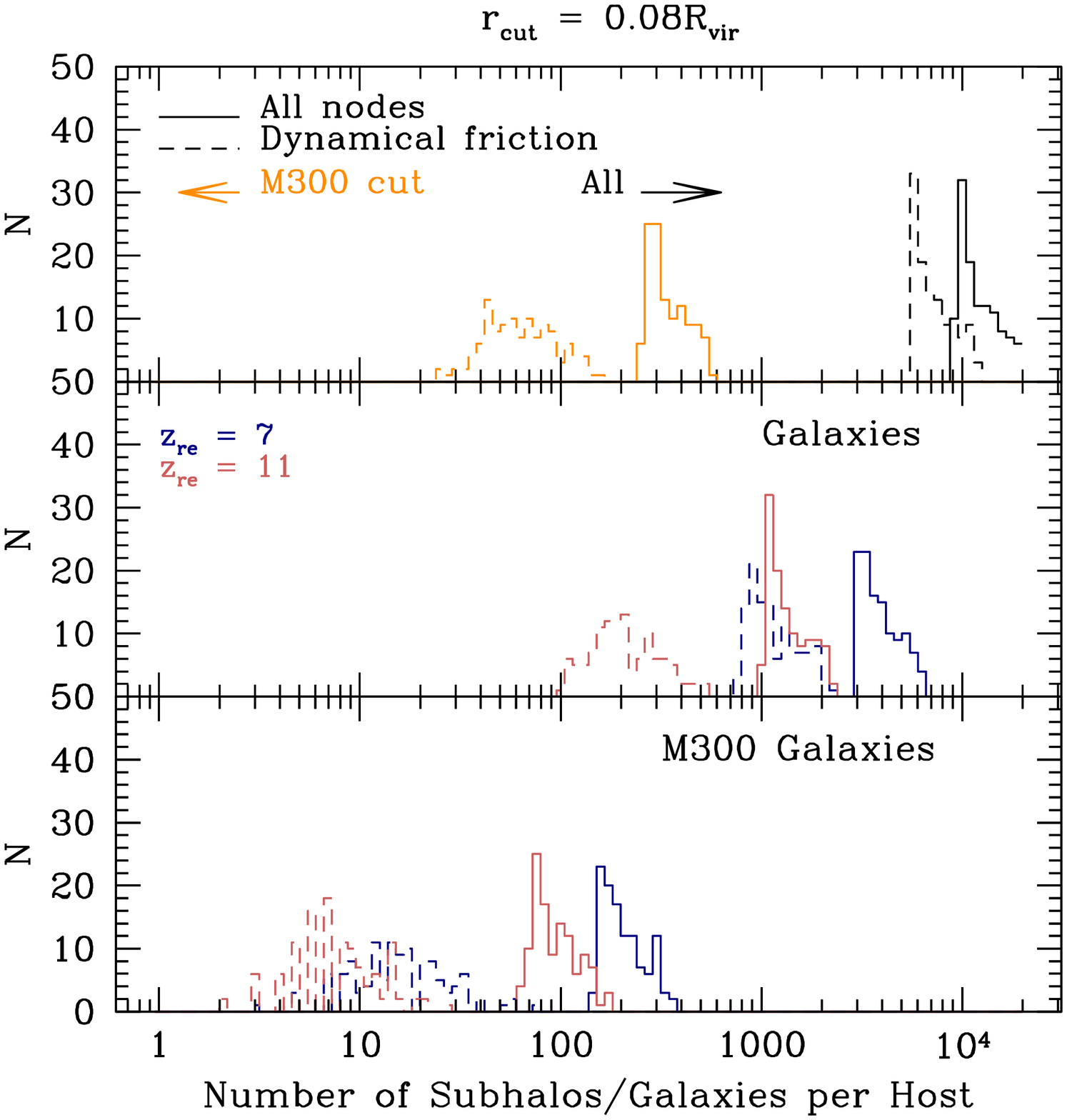}
\caption{\label{fig:m300cut}Dependence of the subhalo/satellite distribution on the inner cut-off for the simulation interpolation table, for $\vke = 30\hbox{ km s}^{-1}$ and $\tau = 20$ Gyr.  \emph{Left:} Inner cut-off well within the numerical relaxation range.  \emph{Center:} Inner cut-off just outside the numerical relaxation range, as determined from simulations in which decay is turned off either at the beginning or later in the simulation.  \emph{Right:} Inner cut-off well outside the numerical relaxation region.  Plot structure identical to that in Fig. \ref{fig:m300comp}.}
\end{figure*}

One potential issue with our hybrid decay simulator is that 300 pc is often lower than the smallest $r/\Rvire$ bin in the halo mass profile (Sec. \ref{sec:methods:simulations}).  This means that we must extrapolate beyond our simulated data to calculate $\mathrm{M}_{300}$.  In general, this means that we tend to \emph{overestimate} $\mathrm{M}_{300}$, since CDM simulations (as well as our decay simulations, in the inner region unaffected by numerical relaxation) find that the inner slope of the density profile tends to become less steep the deeper one gets in the halo \cite{navarro2004}.  We illustrate the effects of the inner radial cut-off of the mass profile on the subhalo and satellite populations in Fig. \ref{fig:m300cut}, with cut-offs of $r = 0.01\Rvire$, $0.04\Rvire$ (default), and $0.08\Rvire$.  For these plots, we set $\vke = 30\hbox{ km s}^{-1}$ and $\tau = 20$ Gyr.  The innermost cut-off, $r = 0.01\Rvire$, is within the numerical relaxation region, in which the density and mass profiles are artificially shallow.  Thus, we tend to find fewer subhalos and satellites that satisfy the $\mathrm{M}_{300}$ and star-formation criteria.  In the rightmost panels, the inner cut-off is set to $r = 0.08\Rvire$.  Here we see that, because the mass profile is a bit steeper here than at $r = 0.04\Rvire$, we tend to overestimate $\mathrm{M}_{300}$, and hence we find that far more satellites and subhalos satisfy the criteria.  Given that the mass profile ought to be becoming shallower inwards of $r= 0.04\Rvire$, we are often still overestimating $\mathrm{M}_{300}$, and hence the constraints on $\vke-\tau$ space based on the work in this section are quite conservative.

\begin{figure*}[t]
\centering
\includegraphics[angle=270,width=0.49\textwidth]{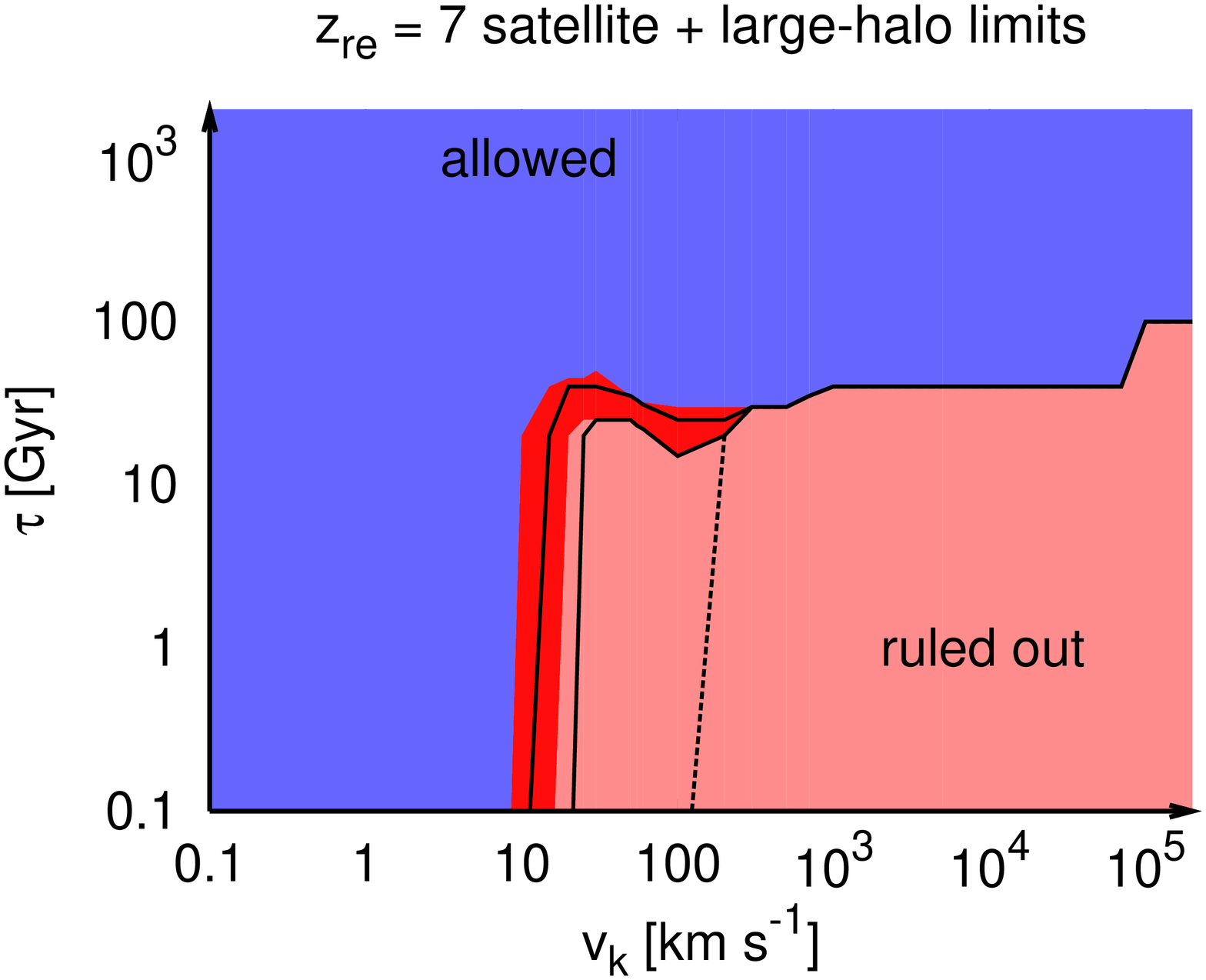}
\includegraphics[angle=270,width=0.49\textwidth]{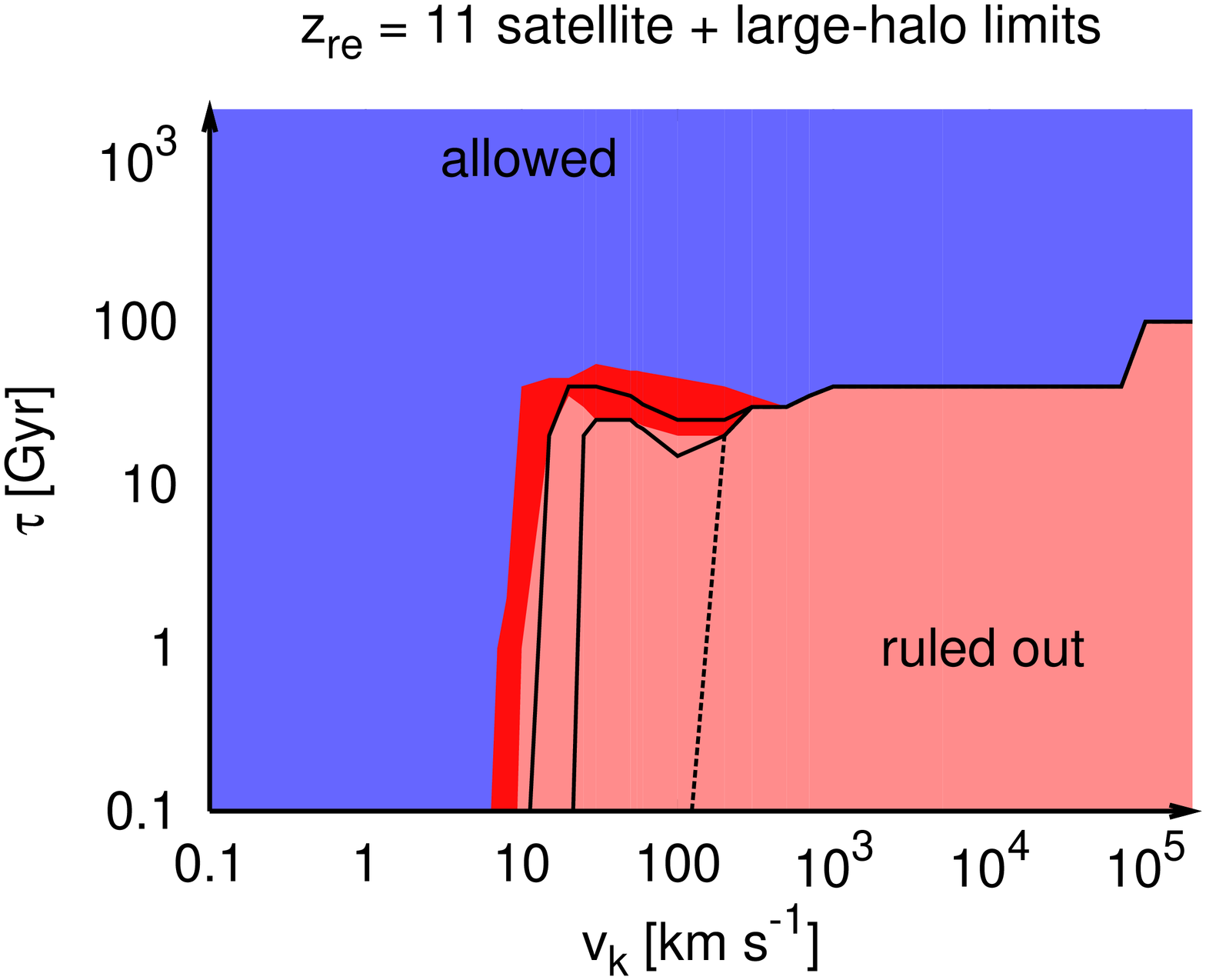}
\caption{\label{fig:exclusion}Exclusion limits in the $\vke-\tau$ parameter space.  In both plots, the blue region marked ``allowed'' indicates the region of parameter space that has not yet been excluded.  The red region marked ``ruled out'' and to the right of the dashed line is the part of parameter space that has been ruled out by observations of the galaxy-cluster mass function and the mass-concentration relation in galaxies, groups, and clusters \cite{ichiki2004,peter2010a,peter2010c}.  The solid lines and the regions to the left of the dashed lines show limits from this work.  The lower solid line shows the limit on the parameter space from the ``all nodes'' subhalo sample, and anything below the upper solid line is also excluded based on the ``dynamical friction'' subhalo sample. The light red region corresponds to limits using the sample of ``all nodes'' subhalos satisfying the star formation criterion with redshift $\zre$, while the dark red region shows the additional excluded region using the ``dynamical friction'' satellite sample satisfying the same star formation criterion.  \emph{Left panel: $\zre=7$}  \emph{Right panel: $\zre = 11$.} }
\end{figure*}

In Fig. \ref{fig:exclusion}, we show our exclusion regions based on the number of star-containing satellites with $\mathrm{M}_{300} > 5\times 10^6\Modot$.  In order to exclude a point in $\vke-\tau$ parameter space, we require that less than 5\% of the host halos in the sample have at least 200 satellites that satisfy both the star-formation and $\mathrm{M}_{300}$ criteria.  A point is allowed if at least 5\% of the hosts have at least 200 satellites.  One can see that all the CDM samples (Fig. \ref{fig:m300cdm}) have more than sufficient subhalos.  

We show constraints for the $\zre = 7$ (left) and $\zre = 11$ samples (right).  The region below the solid lines and to the right of the dashed line was previously excluded by the $z=0$ galaxy-cluster mass function and mass-concentration relation \cite{peter2010c}.  The Milky Way satellite limits are generally less constraining in that region.  The light region to the left of the dashed line corresponds to constraints from the ``all nodes'' samples, and the dark red region corresponds to the additional region excluded by the ``dynamical friction'' sample.  Overlaid on both plots are the exclusion regions based on the subhalo samples (i.e., without the star-formation criterion).  The lower black line corresponds to the ``all nodes'' limit, while the upper line corresponds to the ``dynamical friction'' sample.  We find that star-formation criterion does affect the exclusion regions, although only at the level of a factor of $\sim 2$ in $\tau$ for fixed \vk~unless $\vke \lesssim 20 \hbox{ km s}^{-1}$.

We also apply constraints from the highest-$\mathrm{M}_{300}$ satellites, which are almost entirely the classical dwarf galaxies for which the sample is currently complete.  Of the 11 classical dwarfs, seven have $\mathrm{M}_{300}$ inferred to be $\mathrm{M}_{300} > 10^7\Modot$, one has a smaller $\mathrm{M}_{300}$, while mass modeling is difficult for the remaining three and has so far precluded robust $\mathrm{M}_{300}$ estimates (Large and Small Magellanic Clouds and Sagittarius) \cite{walker2007,strigari_nat2008}.  Since the LMC and SMC have quite large \vmax, and there is a mild correlation of \vmax~and $\mathrm{M}_{300}$, these two also likely have large $\mathrm{M}_{300}$.  Thus, it is possible to constrain \vk~and $\tau$ by determining the number of simulated satellites with $\mathrm{M}_{300} > 10^7\Modot$, and making sure that the classical dwarfs are accounted for.  When we perform this exercise, we find similar constraints as for the number of satellites with $\mathrm{M}_{300} > 5\times 10^6\Modot$ only if $\vke \gtrsim 50\hbox{ km s}^{-1}$.  This is again because $\mathrm{M}_{300}$ is mildy correlated with \vmax, and in order for satellites with high \vmax~to be affected strongly by decays, \vk~needs to be significantly greater than \vvir. However, for $\vke \gtrsim 200\hbox{km s}^{-1}$, constraints from the mass-concentration relation are typically stronger unless the $\zre = 11$ ``dynamical friction'' sample characterizes the satellite population well.

\subsection{$N(>v_{max})$}\label{sec:results:vmax}
A standard way to characterize subhalos and satellites is by their velocity function, as illustrated in Fig. \ref{fig:vhist_cdm} \cite{madau2008b,strigari2010}.  This is often used in lieu of the subhalo or satellite mass because \vmax~is relatively insensitive to the definition of the outer edge of the subhalo or satellite.  Here, we explore the velocity function and the possibility of constraints using the \vmax~function of the observed satellite population.  We emphasize that any constraints we find in this section are highly conservative because we determine \vmax~for the subhalos and satellites in the absence of tidal stripping.  Moreover, we compare the velocity functions in the decay parameter space with the velocity function of known dwarfs, corrected only for SDSS sky coverage (but NOT completeness).

We consider a decay model to be ruled out if it fails to produce a sufficient number of satellites to reproduce the observed velocity function, but consider a model to be allowed if it overshoots the velocity function.  In general, it is much easier to reduce \vmax~(e.g., by tidal stripping) than it is to increase it.

\begin{figure*}
\includegraphics[width=0.49\textwidth]{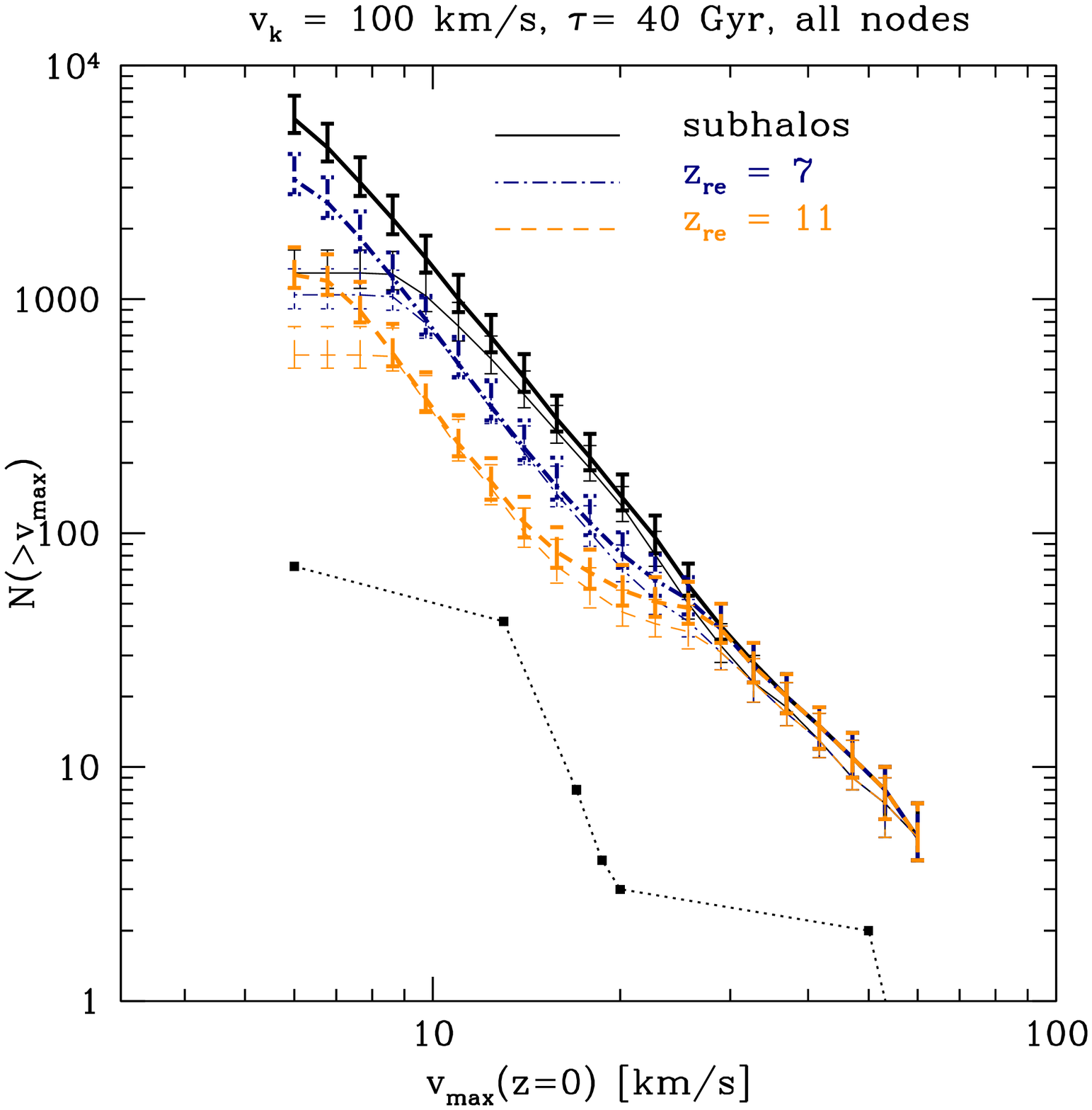}
\includegraphics[width=0.49\textwidth]{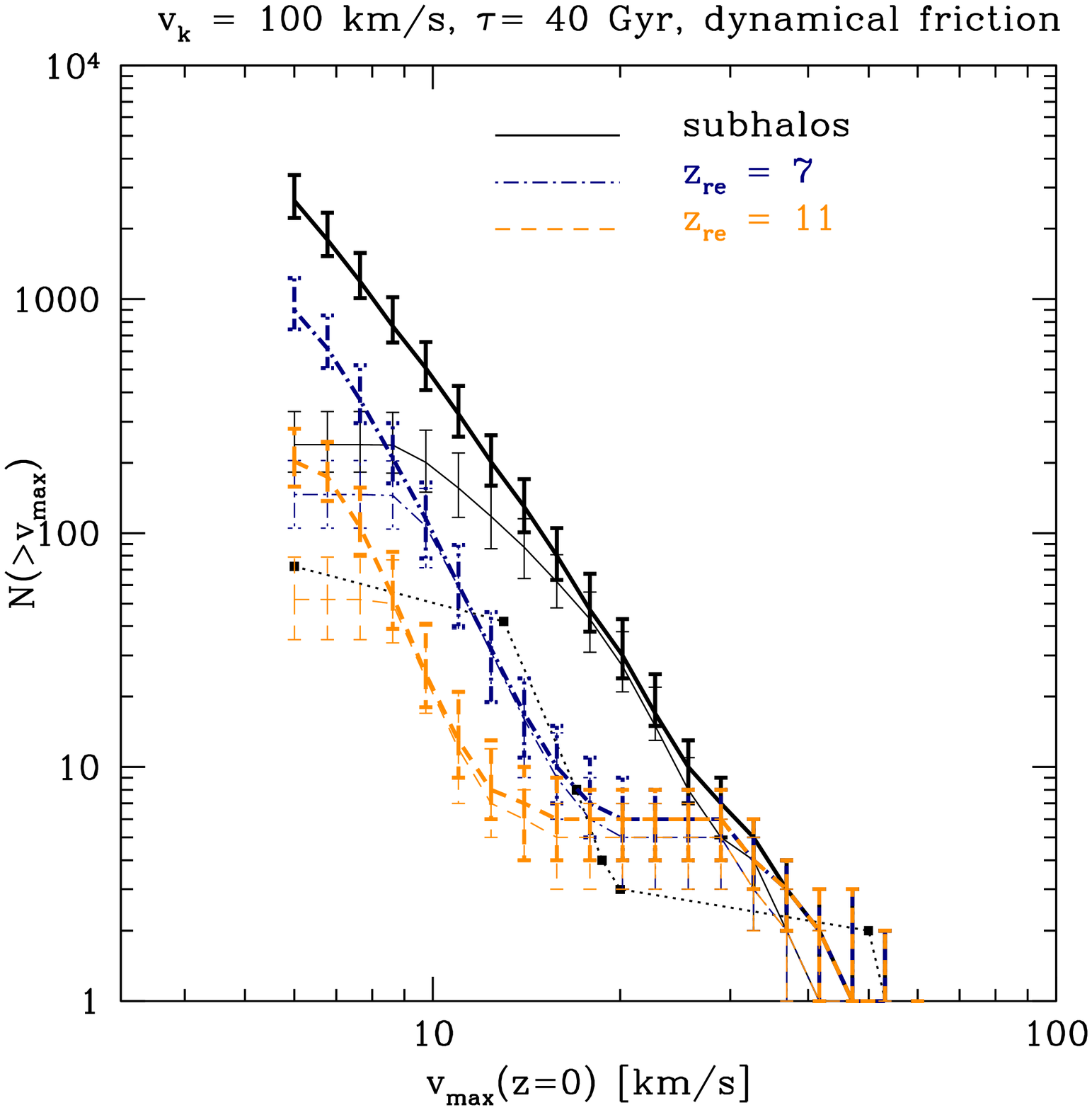}
\caption{\label{fig:vmax_100_40}Maximum circular velocity functions for $\vke = 100 \hbox{ km s}^{-1}$ and $\tau = 40$ Gyr, for ``all nodes'' subsamples (\emph{left}) and ``dynamical friction'' subsamples (\emph{right}).  The line types indicate different subsamples as indicated in the legend: all subhalos, subhalos satisfying the star formation criterion with $\zre = 7$, and subhalos satisfying the star formation criterion with $\zre = 11$.  The thin lines of each type indicate that a cut of $\mathrm{M}_{300}>5\times 10^6\Modot$ has been included, and thick lines show the \vmax~distribution without a cut on $\mathrm{M}_{300}$.}
\end{figure*}

We use Fig. \ref{fig:vmax_100_40} to illustrate a few salient and generic features of the velocity functions.  The velocity function of the ``all nodes'' subhalo population (denoted in Fig. \ref{fig:vmax_100_40} by the thick black line with 25\% and 75\% percentile bars) always lies above the velocity function for the observed dwarf galaxy population corrected for SDSS sky coverage (e.g., the dotted line in Fig. \ref{fig:vhist_cdm} which is that of \citet{madau2008b}, compiled from data in Refs. \cite{mateo1998,munoz2006,martin2007,simon2007}).  With a cut on the subhalo population of $\mathrm{M}_{300} > 5\times 10^{6}\Modot$ (denoted by the thin black line with percentile bars), the velocity function constrains $\tau \gtrsim 20$ Gyr for $\vke \gtrsim 200\hbox{ km s}^{-1}$, which is only competitive with the decay constraint based on the number of subhalos above the $\mathrm{M}_{300}$ threshold for $\vke \gtrsim 100\hbox{ km s}^{-1}$.  At such \vk, it becomes nearly impossible to find subhalos the size of a Large or Small Magellanic Cloud in any Milky-Way mass halo because the decays greatly disturb the large subhalos that would have had large $\mathrm{M}_{300}$ in the absence of decays.  For this range of \vk, though, the observed mass-concentration relation of galaxies rules out a greater swath of $\tau$.  However, for $\tau = 40$ Gyr as illustrated in Fig. \ref{fig:vmax_100_40}, there is no problem forming sufficiently high-\vmax~satellites.

Since, in general, the velocity functions for both the $\zre = 7$ and $\zre = 11$ ``all nodes'' samples merge with the subhalo sample for large \vmax, the constraints on $\vke-\tau$ space are typically identical in the range in which constraints from the velocity function are competitive with those found in Sec. \ref{sec:results:number}.

The velocity function of the ``dynamical friction'' subhalo population produces constraints competitive with the constraints in Sec. \ref{sec:results:number} only for $\vke \gtrsim 200\hbox{ km s}^{-1}$, or $\vke \gtrsim 100\hbox{ km s}^{-1}$ with the cut on $\mathrm{M}_{300}$, which we show in right-hand side of Fig. \ref{fig:vmax_100_40}.  For the subhalo and $\zre = 7$ satellite samples, this constraint again arises from the fact that it is difficult to produce the Magellanic Clouds.  The constraint is tighter for the ``dynamical friction'' sample than for the ``all nodes'' sample because there are far fewer high-\vmax~subhalos even in CDM due to the fact that dynamical friction is more efficient for high-mass (and hence, high-\vmax) subhalos.  

The velocity function provides the strongest constraint on $\tau$ for $\vke > 30\hbox{ km s}^{-1}$ for the $\zre = 11$ ``dynamical friction'' satellite sample, such that $\tau < 60 \hbox{ km s}^{-1}$ is excluded for this \vk~range.  The reason for this is apparent in Fig. \ref{fig:vmax_100_40}.  There is simply a dearth of satellites with $\vmaxe > 10 \hbox{ km s}^{-1}$.  This limit is relatively insensitive to the $\mathrm{M}_{300}$ cut, since a large fraction of the $\zre = 11$ ``dynamical friction'' satellites have large $\mathrm{M}_{300}$.  At fixed $\tau$, the velocity functions look quite similar for $10 \hbox{ km s}^{-1} < \vmaxe < 20\hbox{ km s}^{-1}$ across a broad stretch of \vk~due to the fact that most of the subhalos are in the adiabatic regime of decay in which $\vke \gg \vvire$ and $\tau \gg \tdyne$, where $\tdyne$ is the typical dynamical time of particles in the subhalos.  We show the exclusion limits for $\zre = 11$, including both the constraints from $N(>\vmaxe)$ and the results of Sec. \ref{sec:results:number} in Fig. \ref{fig:exclusion}.  Note that these constraints are even stronger than for the galaxy-cluster mass function and the mass-concentration relation for $\vke \gtrsim 200\hbox{ km s}^{-1}$, but only if the $\zre = 11$ ``dynamical friction'' sample is a good representation of the real satellite population.

\begin{figure}
\includegraphics[angle=270,width=0.49\textwidth]{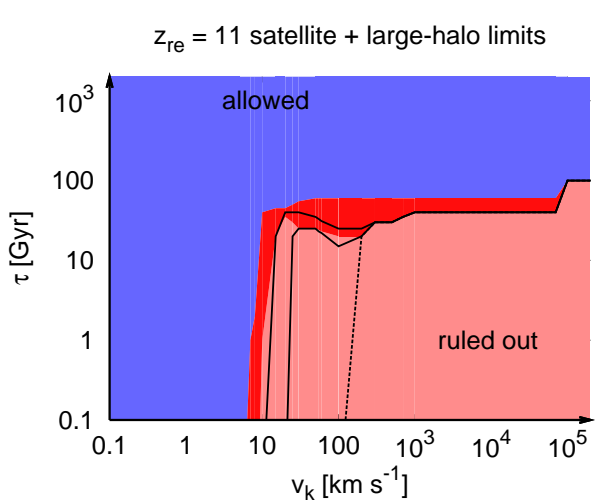}
\caption{\label{fig:vmax_exclusion}Exclusion limits for the $\zre = 11$ satellite populations including the limits from the velocity function.  Lines and shading of the plot have the same meaning as in Fig. \ref{fig:exclusion}.}
\end{figure}

We note that stronger constraints are possible with $N(>\vmaxe)$ if one were to correct the observed velocity function for the SDSS selection function.

\subsection{$\mathrm{M}_{300}$ mass function}\label{sec:results:mf}
\begin{figure}
\includegraphics[width=0.49\textwidth]{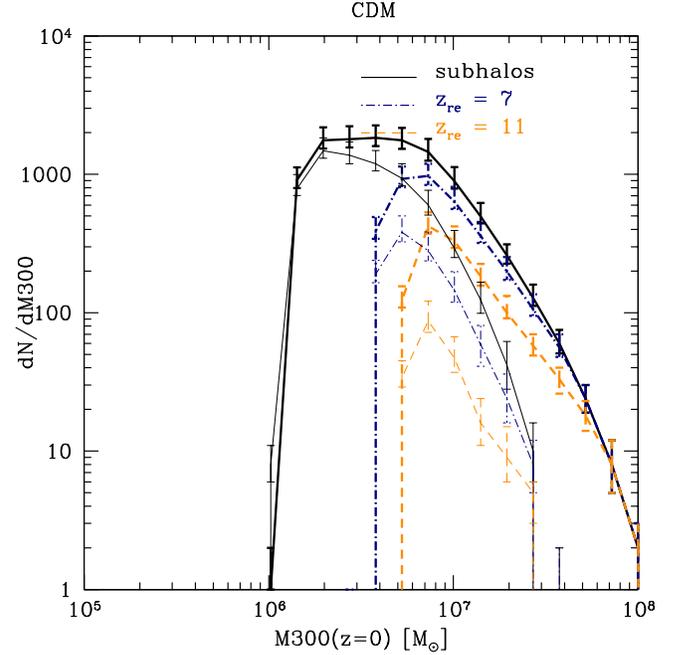}
\caption{\label{fig:dndm300cdm}$\mathrm{M}_{300}$ distribution for CDM.  Line types have the same meaning as in Fig. \ref{fig:vmax_100_40}.  Thick lines represent ``all nodes'' subsamples, and thin lines represent ``dynamical friction'' subsamples.}
\end{figure}

Another possible way to constrain the decay parameter space is to use the full $\mathrm{M}_{300}$ mass function instead of the cuts we employed in Sec. \ref{sec:results:number}.  We show the CDM $\mathrm{M}_{300}$ mass function for our subhalo and satellite samples in Fig. \ref{fig:dndm300cdm}.  We show the mass function for both ``all nodes'' and ``dynamical friction'' subhalo populations, as well as the corresponding satellite populations for $\zre = 7$ and $\zre = 11$.  The cut-off in $\mathrm{M}_{300}$ near $\mathrm{M}_{300} = 10^6\Modot$ for the subhalo populations is an artifact of the mass resolution of the merger trees.  There are several general features of this plot.  First, most of the subhalos and satellites in our samples have $\mathrm{M}_{300}$ in the range corresponding to Milky Way satellites.  Second, the location of the peak of the $\mathrm{M}_{300}$ mass function appears to depend somewhat on the star-formation prescription.  The $\zre = 11$ satellite populations are skewed towards higher $\mathrm{M}_{300}$ than the $\zre = 7$ mass functions regardless of whether we consider the ``all nodes'' or ``dynamical friction'' samples.  In addition, there are far fewer low-$\mathrm{M}_{300}$ satellites in the $\zre = 11$ samples than the $\zre = 7$ samples.  Third, the high-$\mathrm{M}_{300}$ tail depends quite strongly on whether or not dynamical friction is accounted for.  There is a sharp cut-off in all the ``dynamical friction'' subhalo and satellite populations near $\mathrm{M}_{300} \sim 3\times 10^7\Modot$.  This cut-off is only slightly above the observed maximum $\mathrm{M}_{300}$ of the Milky Way satellites for which estimates of $\mathrm{M}_{300}$ exist.  Moreover, the relatively narrow width of the $\mathrm{M}_{300}$ mass functions for the satellite populations suggests that the narrow range of $\mathrm{M}_{300}$ in observed satellites is a \emph{natural} consequence of CDM cosmologies.  The fact that the narrow range of $\mathrm{M}_{300}$ in observed satellites can be simply explained in CDM has been previously noted by \citet{stringer2010}.

\begin{figure*}
\includegraphics[width=0.49\textwidth]{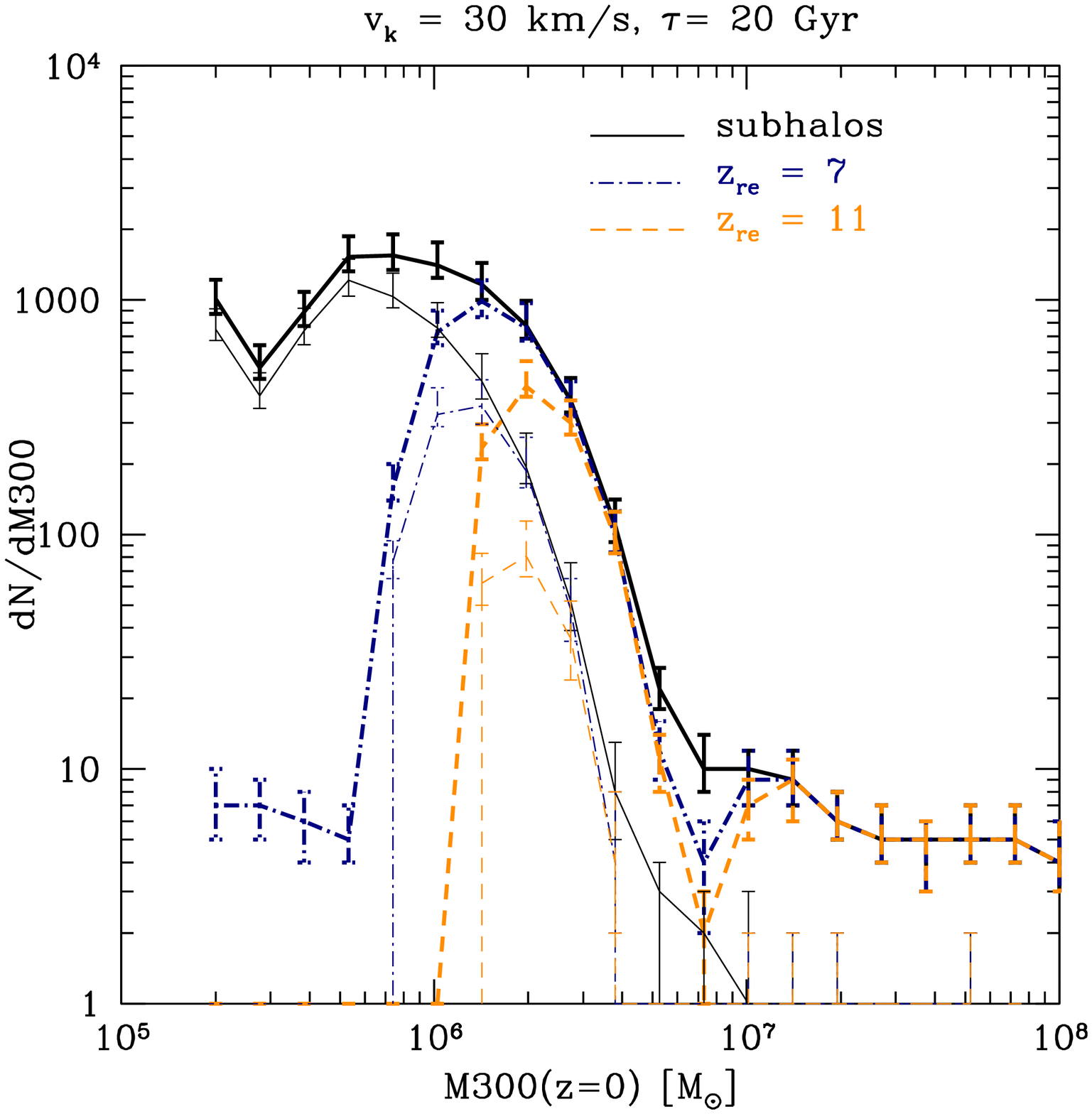}
\includegraphics[width=0.49\textwidth]{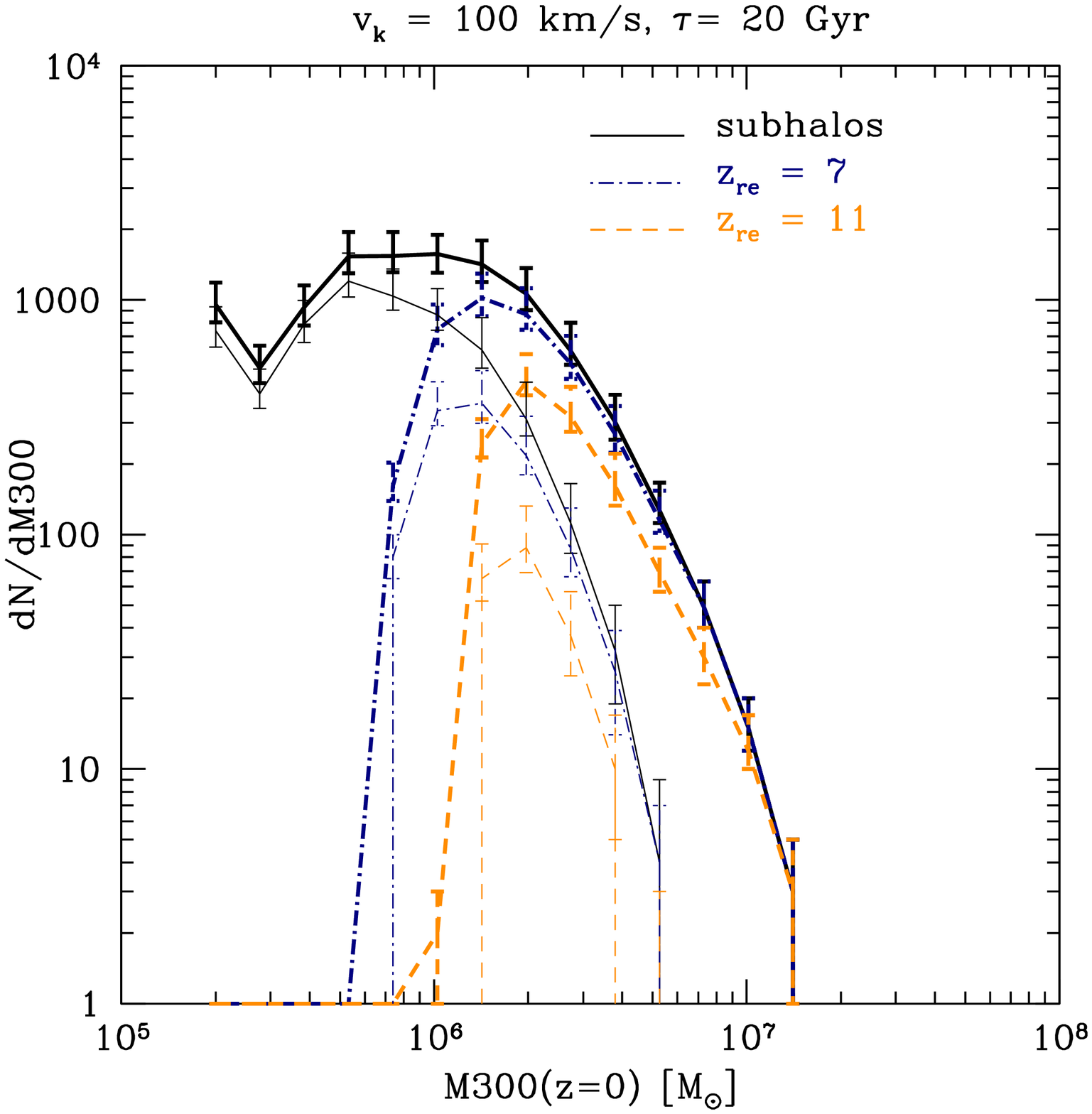}\\
\includegraphics[width=0.49\textwidth]{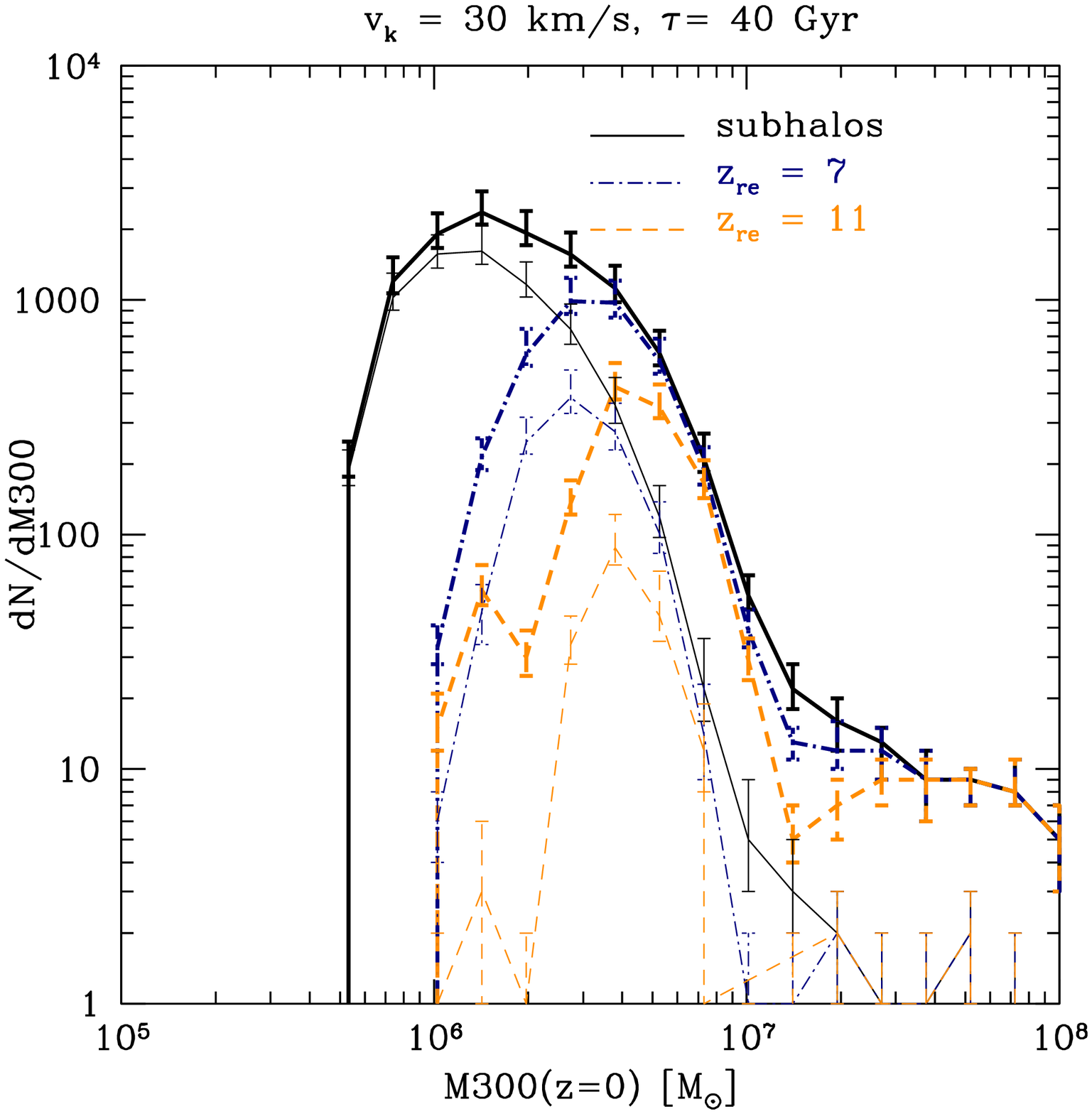}
\includegraphics[width=0.49\textwidth]{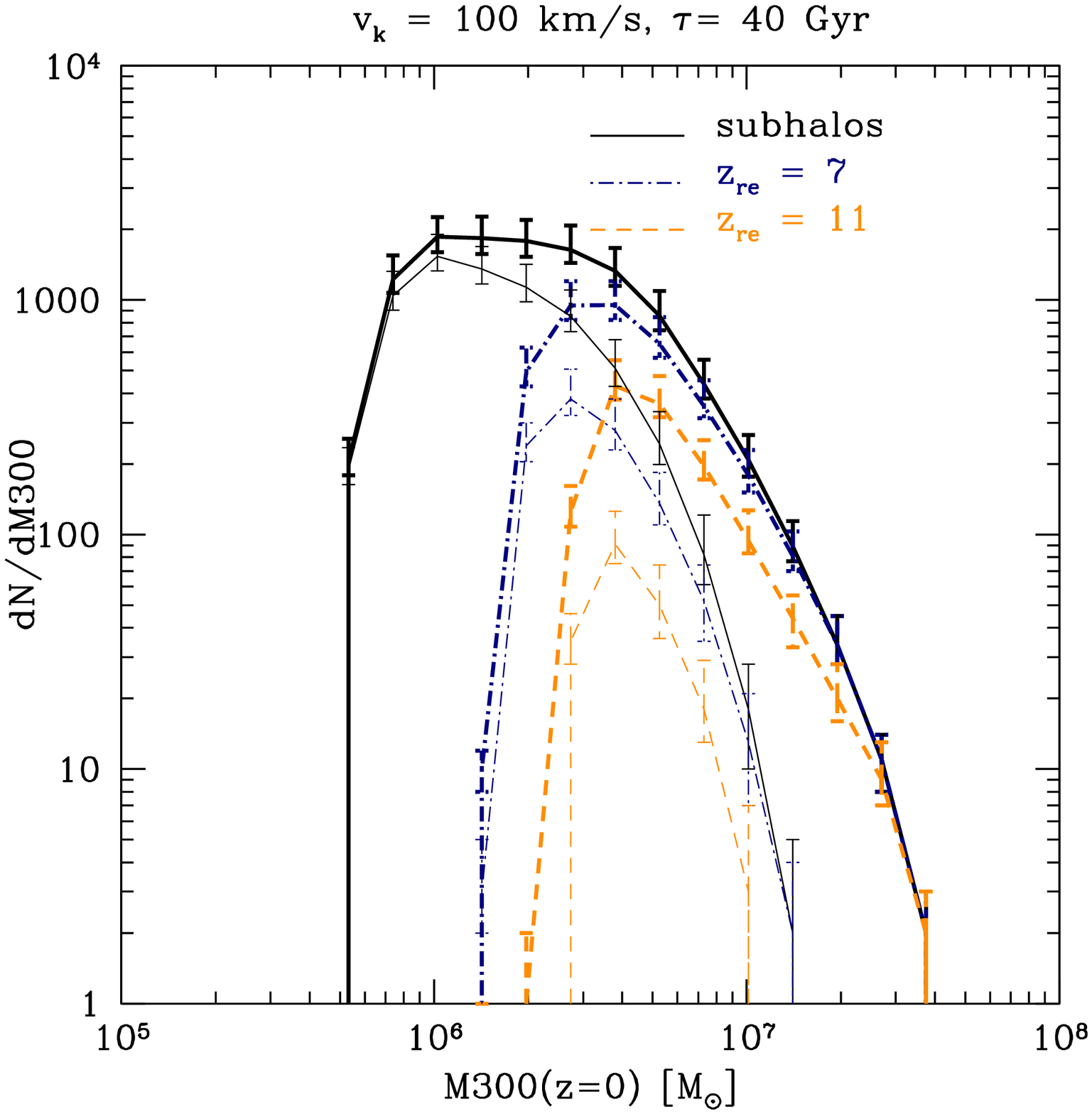}
\caption{\label{fig:dndm300}$\mathrm{M}_{300}$ distributions for those consistent with limits on the $\vke-\tau$ parameter space using the ``all nodes'' subhalo sample (top) and those consistent with the ``dynamical friction'' subhalo sample (bottom).  The line types have the same meaning as in Fig. \ref{fig:vmax_100_40}.  The thick lines represent ``all nodes'' and thin lines represent ``dynamical friction'' subsamples.}
\end{figure*}

We compare the CDM $\mathrm{M}_{300}$ mass function to a few example $\mathrm{M}_{300}$ mass functions for decaying-dark-matter cosmologies, as shown in Fig. \ref{fig:dndm300}.  We show mass functions for $\vke = 30, 100\hbox{ km s}^{-1}$ and $\tau = 20,40$ Gyr.  The mass functions can look quite different, depending on the decay parameters.  In general, smaller lifetimes lead to a much broader smearing of the mass function, with a low-$\mathrm{M}_{300}$ tail predicted for satellite as well as subhalo populations.  While the peak of the mass function necessarily shifts to lower $\mathrm{M}_{300}$ as $\tau$ gets smaller, it is less sensitive to variations in \vk, although the shape of the mass function clearly is quite sensitive.  As \vk~increases, the high-$\mathrm{M}_{300}$ is more sharply cut off since even the largest subhalos begin to become highly disturbed as a result of the decays.

While these plots show that $\mathrm{M}_{300}$ mass functions depend on decay parameters, they also show that they depend on the star-formation prescription and on dynamical processes once the subhalo or satellite is accreted onto a halo.  This is apparent in Fig. \ref{fig:dndm300}, and what is especially striking is the high-$\mathrm{M}_{300}$ tail of the ``all nodes'' samples which extend to far higher $\mathrm{M}_{300}$ than the ``dynamical friction'' or observed satellite samples.  In addition, one effect we have not modeled is tidal stripping.  While 300 pc is typically deep within a satellite or subhalo, the mass within that radius may be reduced as a consequence of tidal stripping, as high-apocenter orbits are progressively removed from the center.  However, if $\mathrm{M}_{300}$ were to be significantly affected by tidal stripping, the associated stellar population should also look significantly disturbed, unless the dark-matter orbits are, on average, highly radial.  Although we hypothesize that the ``dynamical friction'' samples are more likely to represent reality than the ``all nodes'' samples, the shape of the $\mathrm{M}_{300}$ mass function will still depend on the physics of star formation and tidal stripping.  The latter can in principle be modeled using cosmological $N$-body simulations, but would require extremely high resolution. 

There are a few things that are likely to be robust to these effects.  The first is the upper $\mathrm{M}_{300}$ tail of the mass function, since while it is possible to remove mass from inside 300 pc in a halo, it is hard to add mass if there are also only negligible amounts of baryons in the halo (which might have compressed the dark-matter mass profile).  This is essentially what we discussed at the end of Sec. \ref{sec:results:number}, in which we used the classical satellites (which tend to have $\mathrm{M}_{300}\sim 10^7\Modot$) In Fig. \ref{fig:dndm300}, we see that for $\vke = 100\hbox{ km s}^{-1}$ and $\tau = 40$ Gyr, we expect no satellites with $\mathrm{M}_{300} \gtrsim 2\times 10^7\Modot$; if ever such a dense satellite were discovered, it would rule out that point in decay parameter space.

Thus, some aspects of using the $\mathrm{M}_{300}$ mass function to constrain the nature of dark matter are more tractable than others.  A sky- and selection function-corrected $\mathrm{M}_{300}$ distribution function has not yet been published, but there is at least one group working on this (Wolf et al., in prep.).

\section{Discussion}\label{sec:discussion}
In this work, we have shown that the low-\vk~end of the $\vke-\tau$ decay parameter space is currently best constrained by the number of satellites with $\mathrm{M}_{300} > 5\times 10^6\Modot$.  We have found that the precise constraint depends on the star-formation prescription, but not dramatically so (within a factor of 3 for $\tau$ for fixed \vk).  We found that the velocity function of satellites and the $\mathrm{M}_{300}$ values of the classical dwarfs provide similar constraints for $\vke \gtrsim 100\hbox{ km s}^{-1}$, which is because decays only affect the largest subhalos for such \vk.  However, in the case of the $\zre = 11$ ``dynamical friction'' sample, the velocity function more strongly constrains the decay space than the number of satellites above the $\mathrm{M}_{300}$ threshold, the mass-concentration relation, or the galaxy-cluster mass function.  The limits we set in Figs. \ref{fig:exclusion} and \ref{fig:vmax_exclusion} are quite conservative because we use merger trees with high $\sigma_8$ and we use the subhalo properties at accretion to determine \vmax~and $\mathrm{M}_{300}$ at $z=0$.

We showed that the distribution of $\mathrm{M}_{300}$ might be an avenue for future better constraints of decay parameter space, although the distribution does appear to depend on the details of star formation in subhalos.  Conversely, one can think of the $\mathrm{M}_{300}$ distribution as a way to probe star-formation physics as well as dark-matter physics.  In addition, tidal stripping can potentially lower $\mathrm{M}_{300}$, although typically 300 pc is smaller than the radius at which the circular velocity curve of subhalos peaks, making $\mathrm{M}_{300}$ likely a more constant property of a subhalo over its lifetime than \vmax.  Moreover, if $\mathrm{M}_{300}$ were to be significantly affected by tidal stripping, there would likely be evidence of tidal stripping in the stars, too.  There are some generic features in the $\mathrm{M}_{300}$ distributions that likely to be robust, such as the high-$\mathrm{M}_{300}$ tail of the distribution.  

In theory, $\mathrm{M}_{300}$ provides a cleaner and potentially more powerful probe of dark-matter properties than \vmax~because \vmax~typically occurs at larger radii than 300 pc, at least for CDM cosmologies.  The fact that $\mathrm{M}_{300}$ probes the innermost mass of the satellite galaxies is interesting for dark-matter theories that are alternatives to CDM.  All non-CDM theories invoke energy injection or transfer into the dark-matter population by decays (e.g., \cite{dalcanton2001,hogan2000,kaplinghat2005}) or by introducing a non-trivial collision term into the dark-matter Boltzmann equation (e.g., \cite{spergel2000,slatyer2010}).  Typically, the effects of such energy injection or collision has been parametrized by $Q$, the coarse-grained dark-matter distribution function.  $Q$ is enormous for CDM, but becomes is modest once decays or collisions are turned on. 

The problem with trying to infer $Q$ from data is that $Q$ depends on the velocity structure of dark matter, which can never be directly measured (although it may be indirectly measured if, for example, the annihilation cross section is velocity-dependent).  Specific non-CDM theories predict relations between $Q$ and the dark-matter density profile, but one must analyze the data in the context of that specific model \cite{hogan2000,kaplinghat2005}.  While this is useful to constrain specific theories, one cannot generically determine if the observations deviate CDM on the basis of $Q$.

The advantage of using $\mathrm{M}_{300}$ to characterize the dark-matter halos, and using that to consider deviations from CDM, is that any non-CDM dark-matter theory implies that $\mathrm{M}_{300}$ is lower than the CDM value.  Due to the negative heat capacity of self-gravitating halos, kinetic energy injection or collision terms tend to reduce the central density of the halos (at least until core-collapse, in the case of collision terms).  Thus, any physics that would reduce $Q$ would also reduce $\mathrm{M}_{300}$.  Since 300 pc is typically deep within the halo, it could be a good probe of dark-matter properties.

The inferred property of the dark content of Milky Way satellites that has the smallest errors is the mass within the half-light radius of the stars, $\mathrm{M}_{1/2}$ \cite{walker2009,walker2010,wolf2010}.  The half-light radius is often smaller than 300 pc (e.g., $\sim 100$ pc for Coma Berenices \cite{munoz2010,wolf2010}), although it is often larger, especially for the classical dwarfs.  However, the small value of the half-light radius for some of the dwarfs is actually beneficial in constraining the nature of dark matter, since one is measuring the mass within a tiny radius centered on the potential minimum of the halo.

If the $\mathrm{M}_{300}$ mass function is to be a useful probe of either dark-matter physics or galaxy evolution, it is necessary to see how much $\mathrm{M}_{300}$ is affected by the dynamics of subhalos inside the host halos.  However, if this is possible to determine, then in principle, the $\mathrm{M}_{300}$ mass function or the $M_{1/2}$ mass function of satellites should provide an interesting window into dark-matter properties.  Upcoming wide-field surveys, such as the Dark Energy Survey, SkyMapper, and the Large Synoptic Survey Telescope, should reveal many more Milky Way satellites and much more about dark-matter properties \cite{annis2005,keller2007,tollerud2008,lsst2009,bullock2010}.

\begin{acknowledgments}
We thank Manoj Kaplinghat for the discussions that initiated this work, James Bullock and Joe Wolf for stimulating discussions, and Chris Moody and Marc Kamionkowski for some of the simulations used in this work.  We are supported by the Gordon and Betty Moore Foundation.
\end{acknowledgments}

%\bibliographystyle{apsrev}
%\bibliography{dmrefs}

\end{document}